\pdfoutput=1

\documentclass[10pt,letterpaper]{article}
\usepackage[top=0.85in,left=2.75in,footskip=0.75in]{geometry}

\usepackage{amsmath,amssymb}

\usepackage{changepage}

\usepackage[utf8x]{inputenc}

\usepackage{textcomp,marvosym}

\usepackage{cite}

\usepackage{nameref,hyperref}

\usepackage{microtype}
\DisableLigatures[f]{encoding = *, family = * }

\usepackage[table]{xcolor}

\usepackage{array}

\usepackage{pdfpages}

\newcolumntype{+}{!{\vrule width 2pt}}

\newlength\savedwidth

\raggedright
\usepackage[aboveskip=1pt,labelfont=bf,labelsep=period,justification=raggedright,singlelinecheck=off]{caption}

\makeatletter
\renewcommand{\@biblabel}[1]{\quad#1.}
\makeatother

\usepackage{lastpage,fancyhdr,graphicx}
\usepackage{epstopdf}
\fancyheadoffset[L]{2.25in}
\fancyfootoffset[L]{2.25in}
\begin{document}
\vspace*{0.2in}

\begin{flushleft}
{\Large
\textbf\newline{Navigable maps of structural brain networks across species} 
}
\newline
\\
Antoine Allard\textsuperscript{1,2} and
M. \'Angeles Serrano\textsuperscript{3,4,5*}
\\
\bigskip
\textbf{1} D\'epartement de physique, de g\'enie physique et d'optique, Universit\'e Laval, Qu\'ebec (Qu\'ebec), Canada G1V 0A6
\\
\textbf{2} Centre interdisciplinaire de mod\'elisation math\'ematique, Universit\'e Laval, Qu\'ebec (Qu\'ebec), Canada G1V 0A6
\\
\textbf{3} Departament de F\'isica de la Mat\`eria Condensada, Universitat de Barcelona, Mart\'i i Franqu\`es 1, E-08028 Barcelona, Spain
\\
\textbf{4} Universitat de Barcelona Institute of Complex Systems (UBICS), Universitat de Barcelona, Barcelona, Spain
\\
\textbf{5} Instituci\'o Catalana de Recerca i Estudis Avan\c{c}ats (ICREA), Passeig Llu\'is Companys 23, E-08010 Barcelona, Spain
\\
\bigskip

%
%





* marian.serrano@ub.edu

\end{flushleft}
\section*{Abstract}
%
Connectomes are spatially embedded networks whose architecture has been shaped by physical constraints and communication needs throughout evolution. Using a decentralized navigation protocol, we investigate the relationship between the structure of the connectomes of different species and their spatial layout.  As a navigation strategy, we use greedy routing where nearest neighbors, in terms of geometric distance, are visited. We measure the fraction of successful greedy paths and their length as compared to shortest paths in the topology of connectomes. In Euclidean space, we find a striking difference between the navigability properties of mammalian and non-mammalian species, which implies the inability of Euclidean distances to fully explain the structural organization of their connectomes. In contrast, we find that hyperbolic space, the effective geometry of complex networks, provides almost perfectly navigable maps of connectomes for all species, meaning that hyperbolic distances are exceptionally congruent with the structure of connectomes.  Hyperbolic maps therefore offer a quantitative meaningful representation of connectomes that suggests a new cartography of the brain based on the combination of its connectivity with its effective geometry rather than on its anatomy only.  Hyperbolic maps also provide a universal framework to study decentralized communication processes in connectomes of different species and at different scales on an equal footing.

\section*{Author summary}
%
Recent advances in network science include the discovery that complex networks have a hidden geometry and that this geometry is hyperbolic. Studying complex networks through the lens of their effective hyperbolic geometry has led to valuable insights on the organization of a variety of complex systems ranging from the Internet to the metabolism of \emph{E. coli} and humans.  In this paper, we show that this methodology can also be used to infer high-quality maps of connectomes, where brain regions are given coordinates in hyperbolic space such that the closer they are the more likely that they are connected.  Additionally, we find that, even if Euclidean space is typically assumed as the natural geometry of the brain, distances in hyperbolic space offer a more accurate interpretation of the structure of connectomes, which suggests a new perspective for the mapping of the organization of the brain's neuroanatomical regions.


\section*{Introduction}
%
The human brain is arguably one of the most complex systems known to humankind and understanding its inner workings is one of the great scientific challenges of the 21\textsuperscript{st} century~\cite{Alivisatos2012}.  Since the seminal contribution of Santiago Ram\'{o}n y Cajal in the late 19\textsuperscript{th} century, which revealed that brains are at their core networks of discrete individual cells~\cite{ramonycajal}, many efforts have been devoted to uncover the role of the structure and the dynamics of these neural networks in the emergence of cognitive functions~\cite{Bressler2010,Deco2014,Honey2007,Mcintosh1999}.  Building upon the substantial body of work produced over the past century~\cite{Bear2015,Gazzaniga2013,Kandel2012,Bullmore2012}, network science offers a new promising perspective on the brain's networked architecture, known as connectomics~\cite{Hagmann2003,Sporns2005,Sporns2018}.

Current neuroimaging technologies combined with new analytical techniques now allow a systematic extraction of high-resolution neuronal connectivity data in a realistic time~\cite{GrayRoncal2013,Dorkenwald2017}, and hence an increasing number of structural brain networks, or connectomes, are becoming available to the scientific community.  There is, therefore, mounting evidence that most connectomes share \textit{universal} topological properties with other networked complex systems.  For instance, they are modular~\cite{Meunier2010}, small-world~\cite{Bassett2017}, their distribution of number of connections per node is heavy-tailed~\cite{Gastner2016}, and their most connected nodes form a rich-club~\cite{VandenHeuvel2011}.  Interestingly, spatial information about the location of neurons or coarse-grained regions in the Euclidean physical embedding of the brain unveils that geometry plays a part in the organization of connectomes~\cite{Bullmore2012,Ercsey-Ravasz2013,Kaiser2004,Vertes2012}, and might also play a role in communication processes~\cite{Graham2014,Misic2014,Avena-Koenigsberger2017,Seguin2018}. However, Euclidean distances alone cannot explain the observed connectivity between brain regions and other factors are necessarily at play~\cite{Betzel2016,Kaiser2006,Vertes2012}.  This opens the possibility to investigate the relationship between the topology of connectomes and their \textit{effective} geometry~\cite{Ye:2015,Cacciola2017}, i.e. a geometrical representation that encodes all factors guiding the existence of connections.

In this paper, we use an exploratory navigation protocol on connectomes---that works at the interplay between their topology and their effective geometry---to investigate whether distances between brain regions in the embedding space are related with the likelihood of the observed connections. More precisely, we consider connectomes from various species and quantify the efficiency of greedy routing (GR) ---a conceptually simple decentralized navigation protocol~\cite{Kleinberg2006,Boguna2010}--- in Euclidean space (the physical anatomical embedding of brains) and in hyperbolic space (the effective geometry of complex networks in many different domains~\cite{Boguna2010,Serrano2012,Garcia-Perez2016,Alanis-Lobato2018}).  We find a high variability in the efficiency of GR between the connectomes of different species in Euclidean space.  In contrast, their navigability is almost perfect when using maps in the hyperbolic disk. These maps are obtained using techniques in network geometry~\cite{Serrano2008,Krioukov2010}, where connectomes are assumed to exist in an underlying effective geometry that is coupled to the observed topology of connectomes through a universal probabilistic connectivity law that informs about the likelihood of connections between different brain regions. Finally, as reported before for embeddings in hyperbolic space produced by dimensional reduction techniques in machine learning~\cite{Cacciola2017}, we show that our model-based hyperbolic maps reflect known neuroanatomical regions  and functional clusters of the human brain, and so are able to uncover information that was not explicitly incorporated into the embedding of the connectomes.
%
%
%
%
%
\section*{Materials and methods}
%
%
%
\subsection*{Connectome datasets}
%
We use 26 structural connectome datasets covering several species: C. Elegans~\cite{Ahn2006,Varshney2011}, Drosophila~\cite{Takemura2013}, Zebra Finch~\cite{Dorkenwald2017}, Mouse~\cite{Helmstaedter2013,Oh2014}, Rat~\cite{Bota2007}, Cat~\cite{Scannell1995,Scannell1999,DeReus2013}, Macaque~\cite{Harriger2012,Kaiser2006,Markov2014,Markov2013,Young1993} and Human~\cite{Avena-Koenigsberger2014,Fan:2016,GrayRoncal2013,Hagmann2008,Seguin2018}.  Most of the datasets are publicly available from websites like \texttt{icon.colorado.edu} or \texttt{openconnecto.me}, while others were generously shared with us by the authors of the original papers. Out of the 26 connectome datasets, 14 include information on the distances between nodes.  This information is either provided directly via a distance matrix, or indirectly via the position of each region center in 3D Euclidean space.  In the latter case, the distance between two nodes is defined as the length of the straight line between them.

Note that we always used the geodesic distance even though some of the Human datasets also provide the length of white matter fibers between connected regions.  This choice is justified by the fact that the Greedy Routing protocol (see below) also requires the distance between nodes that are not connected.  That being said, we found that geodesic distances and fiber lengths are well correlated, thus making the effect of this approximation on the validity of our conclusions likely to be negligible (see \nameref{tab:fiberlength} for details).  Note also that some connectome datasets are directed.  However, we show on \nameref{fig:directedorundirected} that considering the direction of links only affects marginally the outcome of greedy routing in all but one single connectome (Mouse2), which will be adressed when we analyze the results below.  Consequently, we considered the undirected and unweighted version of each connectome as a common ground for comparison.  See \nameref{app:datasets} for further details on the various datasets.
%
%
%
\subsection*{Greedy Routing}
%
We use greedy routing (GR) as a decentralized navigation protocol to explore the relationship between the topology and the spatial embedding of connectomes. This protocol runs on spatial networks, where source nodes send signals/packets/messages to target nodes. A source node passes the signal along to its neighbor that is the closest to the target node in terms of geometric distance~\cite{Kleinberg2006}. Once the neighbor receives the signal, it repeats the process until the signal either reaches the target (success) or gets stuck in a loop (failure).  When GR succeeds, the path followed by the signal is referred to as a \textit{greedy path}. Most importantly, GR is more likely to succeed if the topological shortest path between two nodes is \textit{congruent} with the path of minimal geometric distance in the underlying space (i.e. the geodesic path, which generalizes the notion of ``straight line'' to curved space), thus motivating the use of the \textit{success rate} to quantify the congruency between a network topology and its embedding space~\cite{Boguna2009a}.  In other words, GR quantifies the extent to which the structure of a network is encoded in the position of its nodes in an underlying geometric space by measuring \textit{how much} do these positions allow to navigate a network efficiently.  Note that greedy paths do not necessarily follow the exact shortest path between two nodes.

\begin{figure}[t]
  \centering
  \includegraphics[width=0.55\linewidth]{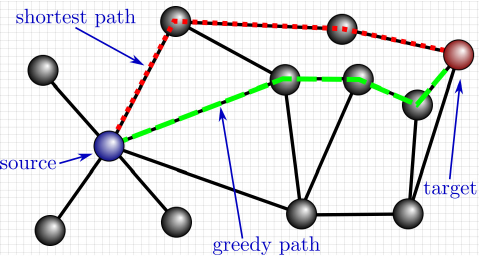}%
  \caption{\label{fig:illustrations}Illustration of a successful greedy path on a network embedded in the 2D-plane (indicated by the grid; distances correspond to the length of the straight line between two nodes).  The corresponding topological and geometrical stretches are respectively 1.33 and approximatively 0.91, thus illustrating how geometrical stretch can be lower than 1.  Notice that the shortest and the greedy paths would coincide if the role of the source and of the target were exchanged (i.e., seeking a greedy path from the red node to the blue one instead).}
\end{figure}

The success rate of GR is computed as the fraction of successful greedy paths when considering every ordered source\slash target pair of nodes belonging to the connected components of a network.  Additionally, greedy paths can be further characterized by their topological stretch, defined as the length in number of links, or \textit{hops} (i.e., topological distance), traversed by the signal when traveling from the source node to the target node, divided by the length of the topological shortest path between them.  A complementary measure is the geometric stretch, defined as the length of the greedy path measured in terms of the sum of the geodesic distances between consecutive nodes on the path (i.e., geometrical distance) divided by the geometrical length of the corresponding topological shortest path.  Note that, contrary to its topological counterpart, the geometrical stretch can take values lower than 1.  Figure~\ref{fig:illustrations} illustrates the various concepts related to GR used in this paper. See \nameref{app:measuring_efficiency_navigation} for a critical assessment of the different options to evaluate the efficiency of greedy routing.
%
%
%
%
%
\subsection*{Hyperbolic embeddings}
%
We use the framework of networks embedded in hidden metric spaces~\cite{Serrano2008,Krioukov2010} to relate the structure of connectomes with their underlying effective geometry.  Connectomes are assumed to exist in an underlying effective geometry that is coupled to their observed topology through a universal probabilistic connectivity law that informs about the likelihood that different brain regions are linked. The positions of the brain regions, or neurons, in the hyperbolic disk are inferred following the procedure described in Ref.~\cite{Boguna2010}.  The hyperbolic maps are based on a purely geometrical model, the $\mathcal{H}^2$ model, in which only the distances between nodes determine their likelihood of being connected~\cite{Krioukov2010}.  More precisely, each node is assigned a radial position, $r$, and an angular position, $\theta$, corresponding to popularity and similarity dimensions respectively. As we will justify below when we introduce the correspondence between the $\mathcal{S}^1$ and the $\mathcal{H}^2$ model, the radial positions of the nodes in the hyperbolic disk is directly related with their hubness. Each pair of nodes, noted $i$ and $j$, is connected with probability
\begin{align} \label{eq:probability_of_connection_H2}
  p_{ij} = \frac{1}{1 + e^{\frac{\beta}{2} (x_{ij} - R)}} \ ,
\end{align}
where $x_{ij}$ is the length of the geodesic between the two nodes in the hyperbolic disk (obtained via the hyperbolic law of cosines), and where the radius of the disk $R$ and the inverse temperature $\beta$ are free parameters fixing the expected average degree and clustering coefficient, respectively.

To understand why the radial and angular coordinates are coined as the popularity and the similarity dimensions, let us consider the isomorphism  given in Ref.~\cite{Krioukov2010} between the $\mathcal{H}^2$ model and the hidden metric space network model $\mathcal{S}^1$~\cite{Serrano2008}, in which the hyperbolic disk is replaced by a one-dimensional unit sphere plus a hidden degree variable associated to each node. In the $\mathcal{S}^1$ model, nodes keep the same angular position as in the $\mathcal{H}^2$ model, but are assigned a hidden degree $\kappa$ that is related to the radial coordinate in the hyperbolic disk according to
\begin{align}
  \kappa \sim e^{(R-r)/2} \ .
\end{align}
Notice that, in real networks, the hidden degree is typically proportional to the observed degree in the topology of the network~\cite{Serrano2008}.  Consequently, nodes closer to the center of the hyperbolic disk have a higher expected degree, and are therefore more \textit{popular}.  Moreover, using the highly accurate approximation $x_{ij} \simeq r_i + r_j + 2 \ln \left( \Delta\theta_{ij} / 2 \right)$ for the length of geodesics in the hyperbolic disk---with $\Delta\theta_{ij}$ being the minimal angular separation between the two nodes---, Eq.~\ref{eq:probability_of_connection_H2} becomes
\begin{align} \label{eq:probability_of_connection_S1}
  p_{ij} = \frac{1}{1 + \left( \frac{\Delta\theta_{ij}}{\mu \kappa_i \kappa_j} \right)^\beta} \ ,
\end{align}
where $\mu$ plays a role analog to $R$.  From Eq.~\ref{eq:probability_of_connection_S1}, we see that nodes are more likely to be connected if they are angularly close, i.e. if they are \textit{similar}, except for high-degree nodes which are likely to be connected regardless of their angular separation (i.e., their ``popularity'' compensates for their lack of ``similarity'').  Note that in these models, the triangle inequality of the underlying metric space---stating that if nodes A and B are both close to node C, then A and B must be close as well---implies a non-vanishing clustering coefficient in the limit $N \rightarrow \infty$, where $N$ is the number of nodes.

From this correspondence, we see that the popularity dimension is related to the degrees of the nodes and is responsible for the hyperbolicity of effective geometry, while the similarity dimension stands as an aggregate of all other factors affecting the likelihood of connections. In other words, in the $\mathcal{S}^1$ model the contribution of degrees is clearly disentangled from other factors reflected in the similarity distance which, for networks like connectomes, may include 3D Euclidean distances among other determinants.

The inferred maps used in this paper are obtained by finding the radial and angular positions of each node, $\{r_i\}$ and $\{\theta_i\}$, that maximize the likelihood that the structure of the connectome has been generated by the $\mathcal{H}^2/\mathcal{S}^1$ model, i.e. maximizing
\begin{align}
  \mathcal{L} = \prod_{i<j} \left[ p_{ij} \right]^{a_{ij}} \left[ 1 - p_{ij} \right]^{1 - a_{ij}} \ ,
\end{align}
where $\{a_{ij}\}$ are the entries of the adjacency matrix of the network ($a_{ij}=a_{ji}=1$ if nodes $i$ and $j$ are connected, $a_{ij}=a_{ji}=0$ otherwise).  This task is achieved using the Metropolis-Hastings algorithm described in Ref.~\cite{Boguna2010}.
%
%
%
\subsection*{Null models}
%
We consider three null models to provide perspective on the success rates of greedy routing in Euclidean and hyperbolic space.  The first null model consists in swapping the positions of the nodes at random (similarly to Ref.~\cite{Kleineberg2016}).  Doing so preserves the spatial distribution of nodes and the topology of the connectome, but destroys its geometry by decoupling the positions of the nodes with its structure.

The second null model preserves the positions of the nodes (i.e. the geometry) but changes the struture of the connectome (i.e. the topology) by swapping pairs of links randomly (also known as the configuration model~\cite{Maslov2002}).  This null model preserves the density, the number of links and the degree sequence of the connectome but destroys correlations in the way nodes are connected as well as destroys modular structure or clustering.  Each randomized version of a connectome was obtained by performing a minimum of $100L$ successful link swaps, where $L$ is the number of links in the connectome.

The third null model is identical to the second one with the additional constraint that each swap is accepted only if it preserves the total cost of the connectome, mesured as the total distance between every connected pairs of nodes, within a tolerance margin (as done in Ref.~\cite{Seguin2018}).  In addition to preserving the density, the number of links and the degree sequence, this procedure preserves the cost of building connections between nodes, of which the total distance between them is a reasonable proxy~\cite{Bullmore2012}.

The first two null models serve as a baseline to be compared with the effect of preserving the total cost in the connectome (third model). To apply the third null model to the wide range of connectomes considered here in a comparable manner, we adapted the procedure described in Ref.~\cite{Seguin2018} in the following way.  Two randomly chosen links, one between nodes A and B and the other one between nodes C and D, are swapped into two new links, one between nodes A and C and the other one between nodes B and D, only if
\begin{align}
  | (d_{AB} + d_{CD}) - (d_{AC} + d_{BD}) | < \varepsilon D \ ,
\end{align}
where $d_{ij}$ is the Euclidean or hyperbolic distance between nodes $i$ and $j$ and where $D = \sum_{i=1}^N\sum_{j=i+1}^N a_{ij} d_{ij}$ is the total distance including every connected pair of nodes (i.e. the total cost).  We set $\varepsilon = 1/60$ to reproduce the criterion used in Ref.~\cite{Seguin2018} for their larger human connectome.  We found that our results are fairly robust and do not depend critically on the precise value of $\varepsilon$.
%
%
%
%
%
\section*{Results}
%
%
%
\subsection*{Navigation in Euclidean space}
%
Before looking into hyperbolic embeddings, we investigate the relationship between the structure of the connectomes and their natural Euclidean embedding using the 14 datasets for which distances are known. We first look at the empirical probability of connection and find that it decreases with Euclidean distance in every dataset, confirming that Euclidean distance is one of the determinants of link formation, as expected~\cite{Bullmore2012,Ercsey-Ravasz2013,Kaiser2004,Vertes2012} (see \nameref{fig:euclidean_connection_probability}).

We then turn to GR (GRE) to further investigate the congruency between Euclidean distances and the structure of the connectomes. Previous work showed that the combination of topology and anatomical geometry in mammalian cortical networks (macaque, mouse, and human brains) allows for near-optimal decentralized communication under greedy routing, in addition to explaining significant variation in functional connectivity~\cite{Seguin2018}.  The effect of rewiring network topology or repositioning network nodes was also showed to cause a 45-60\% reduction in navigation performance, thus suggesting that these brains are naturally configured for near-optimal navigation.  As detailed below, our results confirm and build upon some of the conclusions found in Ref.~\cite{Seguin2018}.

\begin{figure*}[t]
  \centering
  \includegraphics[width = 0.95\linewidth]{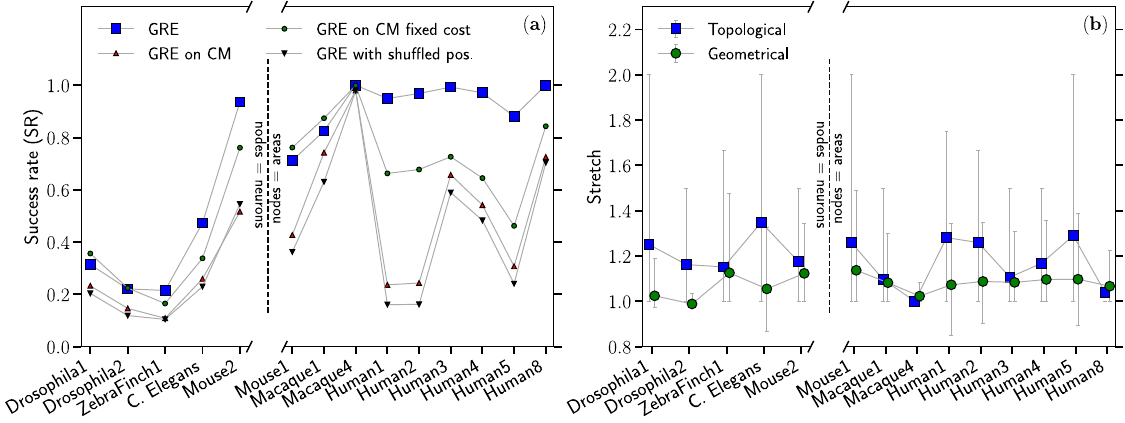}
  \caption{\label{fig:panel_euclidean}
  (a) Success rate (SR) of the greedy routing protocol obtained for connectomes for which Euclidean distance between each pair of nodes in the anatomical embedding is available, as well as for randomized versions of these connectomes generated using the three null models presented in the Materials and Methods section.  The x-axis is broken to highlight the difference between connectomes whose resolution is at the neuron levels (left) from the ones where nodes correspond to coarse-grained regions of the brain (right).  Within these two resolution categories, the ordering of the connectomes from left to right roughly follows the increasing physical volume they occupy.
  (b) Average topological and geometrical stretch of the GRE greedy paths in each connectome with the error bars showing the 10\% and 90\% percentiles.}
\end{figure*}

Figure~\ref{fig:panel_euclidean} shows the success rate and the stretch of GRE where the connectomes have been split based on their resolution (i.e., neuron or coarse-grained regions), and then ordered according to their volume (some datasets, like ZebraFinch1, only correspond to a part of the whole brain).
%
%

Although the number of available connectomes is small, an interesting trend is observed. Since GR quantifies the congruency between the connectome and the underlying geometry, the variability of the success rate shown on Fig.~\ref{fig:panel_euclidean}(a) suggests that data resolution may have a role in the navigability properties of connectomes.  As summarized in \nameref{app:datasets}, the first five datasets on the left on Fig.~\ref{fig:panel_euclidean}(a) are connectomes with a resolution at the neuron level that cover only small volumes of the whole brain (except for C. Elegans which consists in the whole neural system).  These datasets thus miss out on long-distance connections to other regions of the brain, whereas nodes in the remaining connectomes correspond to mesoscopic coarse-grained areas including up to several millions individual neurones.  The low success rates of GRE observed in the connectomes at the neuron-level resolution therefore suggests that their geometry is less congruent with Euclidean geometry.  Note that the high success rate of Mouse2 is likely to be boosted by its unusual large average degree (each node in this connectome is on average connected to about 18\% of all nodes, see \nameref{app:datasets}), which is itself an artefact of the fact that we considered every connectome as undirected. As shown on \nameref{fig:directedorundirected}, this approximation only affects marginally the outcome of greedy routing on every datasets except for Mouse2, where the success rate in the directed version is approx. 50\% lower than in the undirected version. The increased success rate of Mouse 2 in Fig.~\ref{fig:panel_euclidean}(a) could therefore be an artefact of the undirected approximation, thus reinforcing the idea that connectomes are not as navigable at the neuron level than they are at the coarse-grained level. The dependency on the resolution that our results suggest is therefore not a consequence of the use of undirected version of connectomes.

In contrast, coarse-grained connectomes that reflect the large-scale topological organization of connectivity in the brain, and therefore include long-range connections, are marginally affected by density and appear to be more congruent with Euclidean geometry. This last observation is in line with the results obtained in Ref.~\cite{Seguin2018} for mammalian connectomes.  At the same time, the stretch remains very low for all organisms, meaning that the successful greedy paths always follow the shortest paths (topological stretch) closely, and the Euclidean geodesics (geometric stretch) even more closely, as shown in Fig.~\ref{fig:panel_euclidean}(b).

Figure~\ref{fig:panel_euclidean}(a) also shows the average success rate obtained with greedy routing on randomized versions of the connectomes obtained using the three null models presented in the Materials and Methods section.  Our results are in line with the conclusions of Ref.~\cite{Seguin2018} in that cost-preserving topological randomization (third null model) preserves high navigability performances, but can only partially account for the high success rates observed for human connectomes.  Our results are also coherent with the conclusions of Ref.~\cite{Seguin2018} in that cost-preserving topological randomization can account for the high success rates obtained for the macaque (Macaque1 and Macaque4) and the mouse (Mouse1) connectomes and suggest that this conclusion could be extended to the drosophila connectome (Drosophila1 and Drosophila2).  The datasets ZebraFinch1 and C. Elegans stand in-between, with cost-preserving topological randomization preserving a fraction of the low success rates.

\begin{figure*}[t]
  \centering
  \includegraphics[width=0.95\linewidth]{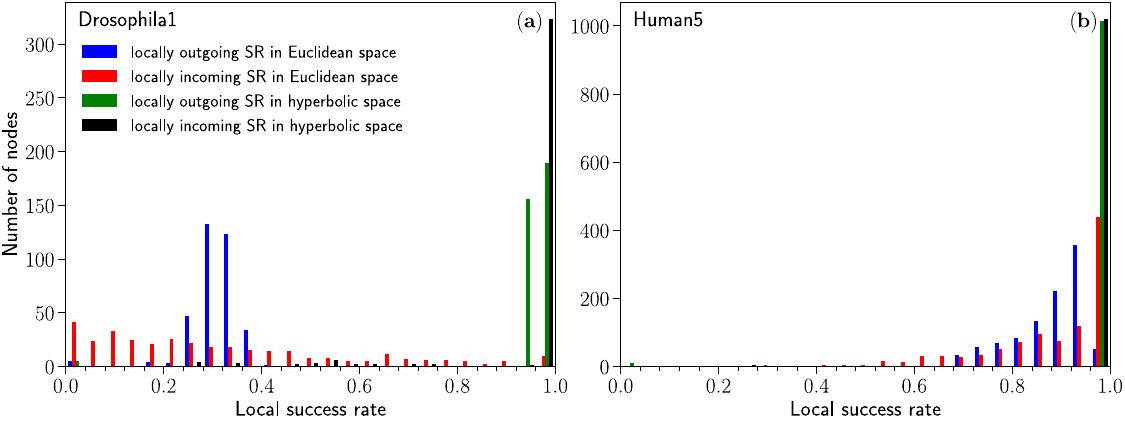}
  \caption{\label{fig:panel_local_sr}
  Distribution of the locally outgoing and incoming success rates for the (a) Drosophila1 and (b) Human5 datasets in Euclidean and hyperbolic space.
  Results for the 14 datasets for which distances in Euclidean space are known are given on \nameref{fig:local_success_rate}.}
\end{figure*}

As a final remark, we noted that the success rate behaves quite differently depending on whether we fix a given node as a source or as a target. If defined as the fraction of successful greedy paths starting from a given source node, we find that these \textit{locally outgoing} success rates are rather tightly distributed around the ``global'' value of SR reported on Fig.~\ref{fig:panel_euclidean}(a).  In other words, greedy paths quickly reach the same set of ``central'' nodes, thus making the success rate only weakly dependent on the identity of the source node.  On the contrary, if defined as the fraction of successful greedy paths aimed at a given target node, we find the distribution of the \textit{locally incoming} success rates to be highly dispersed whenever the global success rate is not close to 1.  This means that, in such cases, the probability of success of a greedy path depends strongly on the target node, and only weakly on the source node. In other words, nodes are not equally reachable: some nodes are more central in Euclidean geometry than others.  Fig.~\ref{fig:panel_local_sr} summarizes these observations using the Drosophila1 and Human5 datasets; results for all 14 datasets are given on \nameref{fig:local_success_rate}.  Note that similar observations were independently reported in Refs.~\cite{Seguin2019b} where the concept of sender/receiver asymmetry is used to characterize brain regions as senders/receivers/neutrals, as well as in Ref.~\cite{Avena-Koenigsberger:2019} where brain regions are characterized based on the information cost when they act as the source or the target of a specific communication path.
%
%
%
%
%
\subsection*{Navigation in hyperbolic maps}
\begin{figure*}[t]
  \centering
  \includegraphics[width=0.9\linewidth]{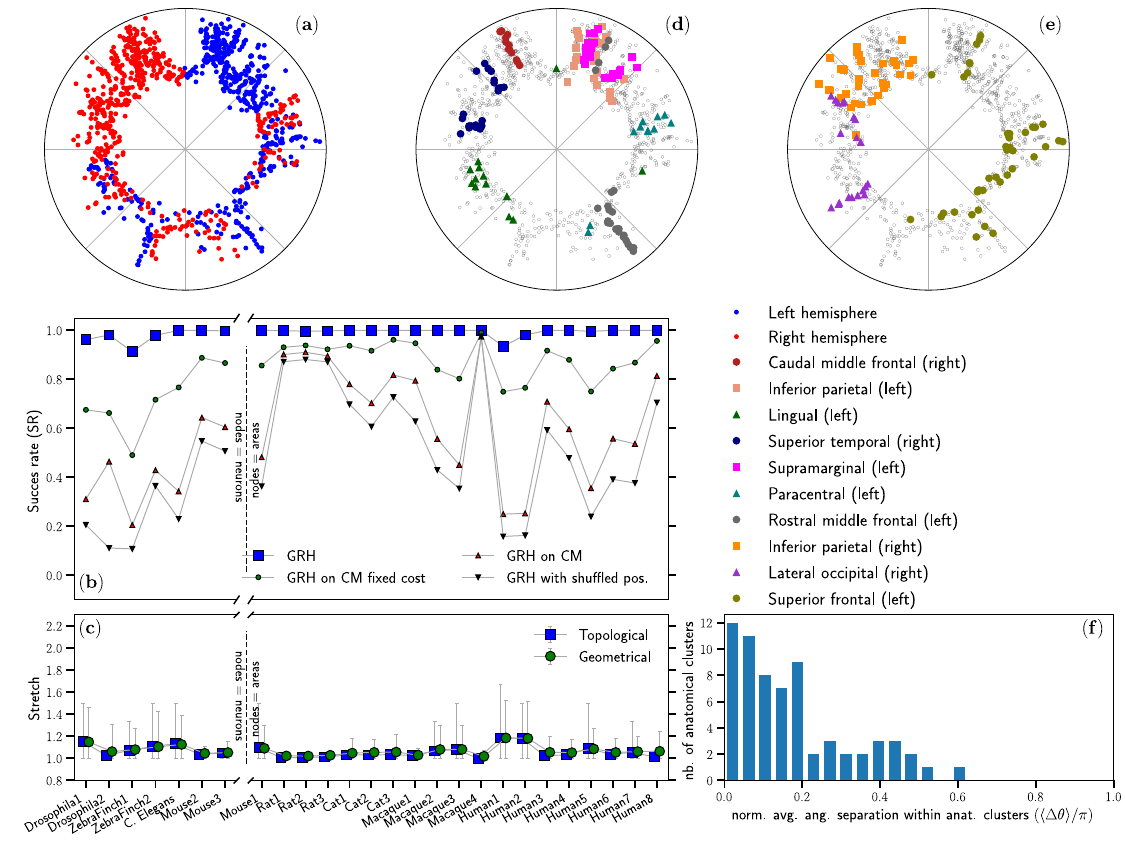}
  \caption{\label{fig:panel_hyperbolic}
  (a) The connectome {\color{black}Human5} embedded in the hyperbolic disk. Nodes belonging to the two different hemispheres are shown in blue and red.  See Materials and Methods for details on this representation.
  {\color{black}(b)} Success rate (SR) of the greedy routing protocol for the hyperbolic embeddings of several connectomes (GRH) as well as for randomized versions of these connectomes generated using the three null models presented in the Materials and Methods section. As in Fig.~\ref{fig:panel_euclidean}, the x-axis is broken to highlight the difference between connectomes at the neuron level from the ones where nodes correspond to areas of the brain.  Within these two resolution categories, the ordering of the connectomes from left to right roughly follows the increasing physical volume they occupy.
  {\color{black}(c)} Average stretch of the greedy paths in each connectome with the error bars showing the 10\% and 90\% percentiles.
  {\color{black}(d)--(e) A sample of representative neuroanatomical regions from the DK atlas are superimposed over the inferred positions of nodes shown on (a).}
  {\color{black}(f) Distribution of the average normalized angular separation between every pair of nodes belonging to the same neuroanatomical regions defined by the DK atlas.}}
\end{figure*}

%
Using the embedding technique described in the Materials and Methods section, we inferred the positions in hyperbolic space of all nodes of the 26 connectomes considered in this study.  Doing so, we obtained a geometrical representation for each connectome without a prior knowledge of the spatial positions of the nodes in Euclidean space.  In the hyperbolic representation, the radial position of a node is directly related to its hubness, with higher degree nodes closer to the center of the disk. Figure~\ref{fig:panel_hyperbolic}(a) shows the hyperbolic map obtained for the Human5 connectome.  Notice that the hyperbolic embedding, which is inferred by using the network topology only, preserves to a large extent the physical separation of the two hemispheres (see also~\cite{Cacciola2017}), with 86\% of the nodes in the left hemisphere (blue) and 82\% of the nodes in the right hemisphere (red) located in separate halves of the disk, even if almost 25\% of the links connect nodes in different hemispheres. This last observation is also clearly reflected on Fig.~\ref{fig:panel_hyperbolic}(a) by the overlap of some regions of the two hemispheres.
%
%

Figure~\ref{fig:panel_hyperbolic}(b) shows that the success rate of greedy routing in the hyperbolic disk (GRH) becomes very close to 100\% for every considered connectome, independently of whether the connectome is at the neuron or coarse-grained region scale.  Moreover, Fig.~\ref{fig:panel_hyperbolic}(c) shows that the greedy paths are very close to their respective shortest paths with average stretches that never exceed 1.2, and less dispersion than in Euclidean space.  These results imply that the network topology of the connectomes is highly congruent with the embedding model $\mathcal{H}^2$ (or equivalently $\mathcal{S}^1$) at all scales; in other words, the inferred coordinates \textit{encode} significant information on the structure of the connectomes, independently of their scale.

Remarkably, reaching these levels of congruency was until now believed to be possible only for networks having a very clean scale-free degree distribution~\cite{Krioukov2010,Boguna2009a}.  As shown in \nameref{fig:complementary_cumulative_degree_distribution}, the degree of the considered connectomes are in general broadly distributed but far from pure power laws. Our results therefore significantly expand the class of networks for which high-quality model-based embeddings in the hyperbolic disk can be obtained. Moreover, the fraction of successful GRH paths does not depend on the specific source or target node, in contrast with results in Euclidean geometry, and remains extremely high for all nodes in all organisms, see Fig.~\ref{fig:panel_local_sr} for representative examples and \nameref{fig:local_success_rate} for all datasets.

Additionnally, Fig.~\ref{fig:panel_hyperbolic}(b) shows that the almost perfect navigability cannot be fully reproduced by neither of the three null models presented in the Materials and Methods section. The only exception is Macaque4, for which density alone is enough to explain the high success rate.  Otherwise, we conclude that reaching such almost perfect navigability requires a near perfect congruency between the topology of the connectomes and their inferred geometry that can hardly be systematically explained by the simple ingredients incorporated in the three null models.  Whatever their exact nature may be, our results show that our embedding method is able to recover these ingredients, independently of the origin of the connectome dataset.

In fact, it is important to stress that these results are obtained even if the embedding procedure is not based on maximizing the coincidence of shortest paths with geodesics in hyperbolic geometry, which would trivially imply a large success rate and small geometrical stretch. In contrast, we maximized the congruency between each connectome and our geometric model based on a connection probability which depends on distances between nodes in the embedding space.  The almost perfect success rates shown on Fig.~\ref{fig:panel_hyperbolic}(b) for GRH therefore suggest a deep, nontrivial relation between the connection probability given by Eq.~\ref{eq:probability_of_connection_H2}, or equivalently by Eq.~\ref{eq:probability_of_connection_S1}, and the structure of the connectomes.  Although the Euclidean distances do encode structural information (see~\nameref{fig:euclidean_connection_probability} and Refs.~\cite{Bullmore2012,Ercsey-Ravasz2013,Kaiser2004,Vertes2012}), Fig.~\ref{fig:panel_euclidean}(a) shows that this information is not enough to explain brain connectivity and varies from one connectome dataset to another.  Interestingly, Euclidean distance cannot be used as a direct estimation of the angular separation, $\Delta\theta_{ij}$ in Eq.~\ref{eq:probability_of_connection_S1}, which characterizes the similarity between nodes.  This observation is further supported by the fact that the correlation between distances in Euclidean space and inferred angular distances in the hyperbolic embeddings never exceeds 0.6 (see \nameref{tab:pearsons}), implying that angular distances encoding similarity include more information related to the structure of connectomes than Euclidean distances.
%
%
%
\subsection*{Meaningfulness of hyperbolic embedding of brain networks}
%
%
\begin{figure*}[t]
  \centering
  \includegraphics[width=0.9\linewidth]{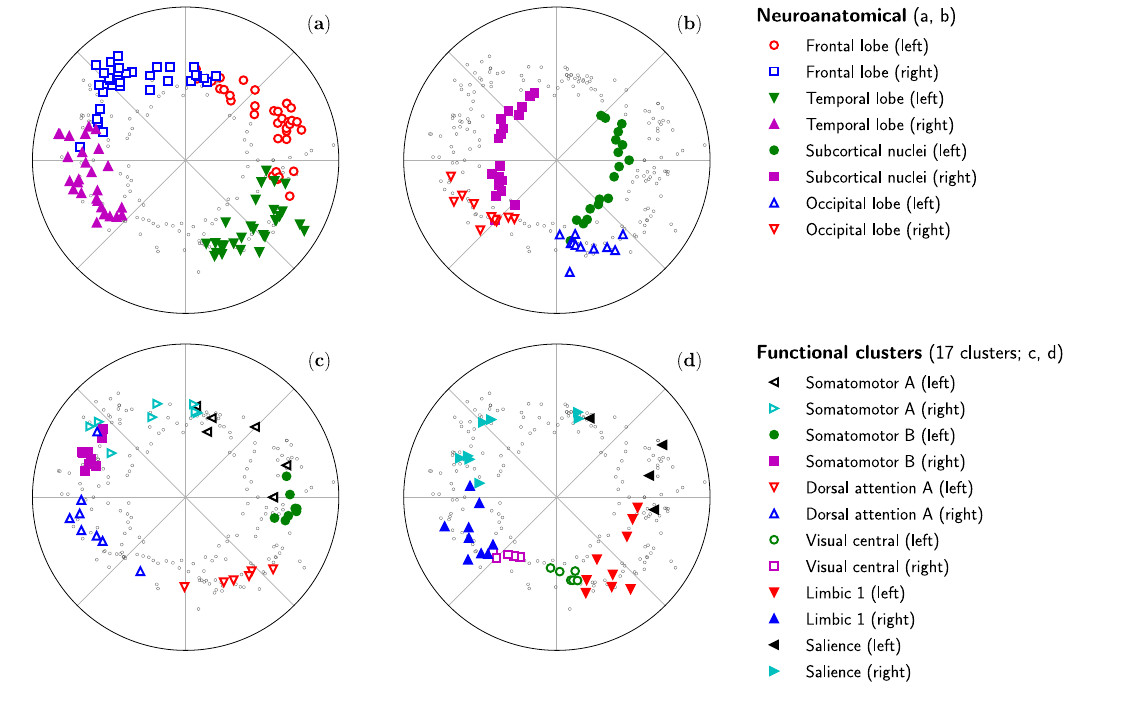}
  \caption{\label{fig:panel_functional}
  Superposition of various neuroanatomical regions and functional clusters on the inferred positions of the nodes in the hyperbolic disk for the Human8 dataset.
  (a--b) The neuroanatomical regions correspond to the lobes as identified by Ref.~\cite{Fan:2016}.
  (c--d) The functional clusters correspond to the 17-region parcellation proposed in Ref.~\cite{Yeo2011} using the name introduced in Ref.~\cite{Baker2014}.
  The clusters shown were chosen so that many could fit on a single plot without clutering it; the complete set of neuroanatomical regions and functional clusters are shown on Fig.~\nameref{fig:anatomical_clusters_angular_positions_Human8}.}
\end{figure*}
As explained above, in the hyperbolic representation the radial position of a node is directly related to its hubness, with higher degree nodes situated closer to the center of the disk. To further explore the structural information unveiled by our hyperbolic maps, we investigate the neuroanatomical relevance of the similarity dimension.  Figures~\ref{fig:panel_hyperbolic}(d)--(e) overlay the positions inferred for the Human5 dataset in the hyperbolic disk with neuroanatomical regions defined by the DK atlas~\cite{Desikan2006}.  As shown on Fig.~\ref{fig:panel_hyperbolic}(f), most neuroanatomical regions of Human5 are localized in narrow regions of the similarity space as measured by the angular separation $\Delta\theta$ in the hyperbolic maps.  We see that nodes belonging to the same neuroanatomical region tend to be strongly concentrated in a narrow angular section, with $84\%$ of the neuroanatomical regions spanning over less than $20\%$ of the similarity space.  This reflects that their nodes are densely connected and have many common neighbors. Interestingly, many of these localized regions spread along the radial coordinate, meaning that their internal structure is dominated by a stratification of degrees and is strongly hierarchical~\cite{Garcia-Perez2016}.  Contrarily to this trend, a few neuroanatomical regions display a dispersed spectrum of angular positions (e.g., rostral middle frontal), denoting more interconnections with other regions and therefore a more prominent role in providing integration at the system level.  Similar conclusions can be made for the Human8 dataset on Fig.~\ref{fig:panel_functional}(a)--(b).  The complete comparison between neuroanatomical regions and the inferred hyperbolic maps for the full Human1, Human2, Human3, Human5 and Human8 datasets is provided on \nameref{fig:anatomical_clusters_angular_positions_Human5} to \nameref{fig:anatomical_clusters_angular_positions_Human8}.  In addition to a visual appreciation of the localization of the neuroanatomical regions in the hyperbolic disk, these figures provide the angular distribution of each region, as well as the likelihood of finding clusters at least as angularly localized as the original ones by randomly selecting the same number of nodes.  We found that in far less than 1\% of the 10000 random samples generated for 52 neuroanatomical regions in Human5 (out of 68) a random selection of nodes resulted in an average angular distance between them smaller than the one measured for the original regions.
Similar results were obtained for the Human1, Human2, Human3 and Human8 datasets for which the localization of 22/32, 31/32, 30/52 and 52/64 of the neuroanatomical regions, respectively, could not be explained by chance.  The congruency between neuroanatomical regions and hyperbolic embeddings observed here is in line with the conclusions of Ref.~\cite{Cacciola2017} where a similar matter was discussed.

The inferred angular positions also offer information about the community structure of the analyzed connectomes. We compared the communities identified by the critical gap method (CGM, see \nameref{app:critical_gap_method}) \cite{Serrano2012,Garcia-Perez2016} with the neuroanatomical regions available for the Human1, Human2, Human3, Human5 and Human8 datasets. We found that the inferred communities and neuroanatomical regions overlap significantly, with normalized mutual information (NMI) values between 0.42 and 0.54 for all five datasets.  On the one hand, this further confirms that the inferred angular positions contain meaningful information related to the known neuroanatomical regions.  Interestingly, this result also suggests that a new cartography of the brain's regions that combines connectivity and effective geometry (e.g., inferred angular positions of nodes) could complement other existing classifications~\cite{Genon2018}.

Beyond neuroanatomical regions, the hyperbolic maps reveal also information related to functional communities.  Figs.~\ref{fig:panel_functional}(c)--(d) superimpose the inferred positions of the nodes in the Human8 connectome datasets with several clusters of the 17-cluster parcellation proposed by Yeo \textit{et al.}~\cite{Yeo2011}. The clusters shown were selected to display simultaneously the maximum number of clusters while avoiding information overload. Results for all the functional modules in the full 7-cluster and 17-cluster parcellations of Ref.~\cite{Yeo2011} are given on \nameref{fig:anatomical_clusters_angular_positions_Human8}. A priori, the localization of functional clusters in the hyperbolic embedding of the connectome was not obvious. However, we found that 14 out of 14 regions of the 7-cluster parcellation (each hemisphere was considered individually) and 22 out of the 33 regions of the 17-cluster parcellation (1 region was excluded because it only has one node in the left hemisphere) were localized angularly in a way that could not be explained by chance (we used the same procedure as for neuroanatomical regions to estimate the average angular distance in the random case). Quantitative results are reported in \nameref{fig:anatomical_clusters_angular_positions_Human8}.

Altogether, these results suggest that the hyperbolic distance obtained via our embedding procedure offers a meaningful effective distance as an aggregated measure of all factors that determine the likelihood of connections.  Indeed, \nameref{fig:hyp_dist_vs_euclid_dist} shows that Euclidean and hyperbolic distances tend to be correlated, but the striking difference between the success rates of GRE and GRH implies that distances in hyperbolic space are not a mere translation of the distances in Euclidean space.  As a corollary, the geometrical representation in hyperbolic space, as in Fig.~\ref{fig:panel_hyperbolic}(a), offers a quantitatively meaningful way to visualize and compare the structure of connectomes.
%
%
%
%
\section*{Discussion}
%
Many real networks are naturally embedded in a physical space that contributes to shape their structure and organization. This is the case of connectomes, whose architecture has evolved in 3D Euclidean space to perform different functions.  However, it is not clear to which extent the anatomical spatial layout of brain regions is informative of how they are connected. We investigated this question by computing the success rate and stretch of greedy routing, a decentralized routing protocol typically used to explore the navigability properties of networks.

Our results suggest a dependence of the success rate of GRE on the resolution of the connectomes.  We find that neuron-level connectomes (e.g., Drosophila1 and ZebraFinch1) are significantly less congruent with Euclidean geometry than connectomes in which nodes correspond to coarse-grained regions of the brain (e.g., Macaque1, Human8).  This difference could be due to the fact that neuron-level connectomes only cover small volumes of the brain, and therefore lack long-distance connections to other regions of the brain, whereas coarse-grained connectomes cover most of the brain and therefore capture these long-range connections.  Interestingly, our results also raise the question as to whether perfect navigability should be expected at the cellular scale since every neuron may not need to communicate with every neuron.  It will be possible to further explore this hypothesis in the near future as more comprehensive and more detailed connectomes become available.

Beyond Euclidean space, we showed that all the considered connectome datasets are strikingly congruent with our model in hyperbolic geometry.  These results suggest that a universal connectivity law can successfully describe the topology of connectomes of different species and at very different scales (e.g., there are several order of magnitude of resolution between Drosphila1 and Macaque1).

An interesting fact about navigation in hyperbolic space is that it naturally reproduces the ``path motifs'' identified in Ref.\cite{VandenHeuvel2012} in which shortest communication paths pass through nodes of increasing and then decreasing degree (i.e., pass through the rich-club, see Ref.~\cite{Boguna2009a}). The almost-perfect success rates observed on Fig.~\ref{fig:panel_hyperbolic}(b) therefore mean that greedy routing was able to recover most of these shortest communication paths, and consequently that the hyperbolic maps that we obtained for the 26 datasets naturally encode the information to simulate the way information flows in the brain.

Moreover, previous studies showed that the topology of connectomes is better reproduced when distances in Euclidean space are combined with topological information such as degrees, clustering coefficients or common neighbors (homophily)~\cite{Betzel2016,Vertes2012}.  Critically, this information is naturally taken into account in our approach via Eq.~\eqref{eq:probability_of_connection_S1} (degrees) and via the triangle inequality of the hyperbolic space (clustering and common neighbors).  Our hyperbolic framework therefore emcompasses the pivotal points of previous approaches proposed to model the topology of connectomes.  Most importantly, our approach allows a richer description than most models proposed in the literature since similarity distances are not limited to the Euclidean distances and can account for other factors influencing the existence of connections.  In fact, our results support hyperbolic geometry as a meaningful \textit{effective} geometry of connectomes.

Indeed, just as in other contexts~\cite{Boguna2010,Serrano2012,Garcia-Perez2016}, hyperbolic maps not only explain the large-scale connectivity of brains but also encode relevant information about their mesoscale organization that is not explicitly annotated into the embedded topology.  These maps therefore offer a complementary and meaningful geometrical representation of connectomes for which the Euclidean physical positions of the nodes need not be known.  We illustrated this point by overlaying known neuroanatomical regions and functional clusters with our hyperbolic embeddings for human connectomes and found an impressive congruency.  Similarly to biochemical pathways in metabolic networks~\cite{Serrano2012}, many brain regions are localized in similarity space with a radial stratification, which implies a modular and hierarchical architecture.  In contrast, a few regions are more angularly dispersed, which indicates a higher density of inter-regional links and therefore a fundamental role in providing system-level integration.  Altogether, this means that the positions of nodes in the hyperbolic embeddings carry meaningful information about the overall organization of brains.  This new geometric information---which could be based on a spatial clustering of nodes in the hyperbolic disk to combine information about both the connectivity of nodes and their similarity---could therefore be used to enrich newly proposed parcellation methods of the human brain based on connectivity~\cite{Arslan2018,Fan:2016}.

Little is known about large-scale communication processes in the brain~\cite{Graham2014,Misic2014,Avena-Koenigsberger2017,Seguin2018} and a key concern for graph-theoretic neural signaling models---of which GR is a very simple example---is the quest for biological plausibility~\cite{Avena-Koenigsberger2017}.  On the one hand, shortest path routing relies on the unrealistic assumption that neural elements possess global knowledge of the network topology while, on the other hand, random diffusion needs bias to travel via efficient routes in realistic times.  GR proposes a compromise scenario in which only the geometrical positions of first neighbors and of the target node are needed at each step of the routing process.  Although there is some evidence that targeted information processing may play a role in brain communication dynamics~\cite{Ciocchi:2015}, an empirical justification that regions/neurons could possess knowledge on the spatial positioning of their neighbors in relation to a target region/neuron is still lacking (see Ref.~\cite{Seguin2018} for a detailed discussion).  In hyperbolic space, nodes are endowed with even more information, since distances in hyperbolic space include other factors than merely anatomical distance.  It may therefore not come at a surprise that navigability is higher in hyperbolic space.  However, independently of whether GR is a realistic neural communication process or not, our results imply that hyperbolic space can quantitively and meaningfully encode the structure of connectomes, and therefore that network geometry offers a valuable complementary approach to study the structure of the brain.

Future work will require the adaptation of the network geometry toolbox to increase the number of realistic features that can be taken into account.  For instance, effective distances may be affected by dynamical ingredients that influence the speed of synaptic transmission and even be modulated in a task-dependent fashion~\cite{Deco:2011,Haimovici:2016,Shimono:2018}. In particular, myelination of neural fibers can dramatically change this speed, especially for long white-matter fibers, thereby effectively shortening Euclidean physical distances.  It would therefore be interesting to analyze GR as a routing protocol in hyperbolic maps of connectomes that are able to modulate effective distances according to various factors.  Additionally, many structural brain networks are characterized by a heterogeneous distribution of fiber densities.  A natural extension of our work would therefore be to study the effects of weights in neural communication for which new methodologies like a technique to embed weighted networks in hyperbolic space are still under development.

This study puts forward an interesting new path to further explore the synergy between neuroscience and network geometry to better understand the inner workings of the brain.  In the future, higher resolution datasets will allow to refine the scope of our conclusions, further test the formulated hypotheses and, ultimately, lead to a better understanding of the interplay between the organization of the connectome and its embedding space.

\section*{Supporting information}

\paragraph*{S1 Table}
\label{tab:fiberlength}
Pearson and Spearman's rank correlation coefficient between the fiber lengths provided in the Human datasets and the corresponding geodesic lengths used for Greedy Routing. Also shown are estimations of the 95\% confidence interval obtained by 10000 resampling of the original data.

\paragraph*{S2 Table}
\label{tab:pearsons}
Pearson correlation coefficient of the distances between nodes in the 3D Euclidean space and the inferred angular distances in the hyperbolic embeddings.

\paragraph*{S1 Appendix}
\label{app:datasets}
Overview of the different datasets.

\paragraph*{S2 Appendix}
\label{app:measuring_efficiency_navigation}
Critical assessment of the different options to evaluate the efficiency of greedy routing.

\paragraph*{S3 Appendix}
\label{app:critical_gap_method}
Critical Gap Method.

\paragraph*{S1 Fig.}
\label{fig:euclidean_connection_probability}
Connection probability in Euclidean geometry for every connectom datasets for which spatial information is available.  The connection probability is computed as the fraction of node pairs that are connected as a function of the Euclidean distance between them. The distances have been rescaled (``normalized'') to facilitate the presentation of the curves.

\paragraph*{S2 Fig.}
\label{fig:local_success_rate}
Distribution of the local success rates defined as the fraction of successful greedy paths starting/aiming at a given source/target node $i$, for all nodes. These distributions are shown for Greedy Routing in Euclidean space (blue and red) as well as in the hyperbolic embeddings (green and black).

\paragraph*{S3 Fig.}
\label{fig:complementary_cumulative_degree_distribution}
Complementary cummulative degree distribution of the connectome datasets.

\paragraph*{S4 Fig.}
\label{fig:hyp_dist_vs_euclid_dist}
Distances between nodes in Euclidean space compared to the inferred distance between the same nodes in the hyperbolic embedding for the connectome datasets for which spatial information is provided.  Distances have been rescaled to facilitate the presentation.  Also shown is the Pearson correlation coefficient, $\rho$, between the two sets of distances. Note that the horizontal position of some markers have been slightly offset to reduce the overlap of the error bars.

\paragraph*{S5 Fig.}
\label{fig:directedorundirected}
Comparison of the success rate of greedy routing in Euclidean space for the directed and undirected versions of the 14 connectome datasets for which distances/positions in Euclidean space are available.

\paragraph*{S6 Fig.}
\label{fig:anatomical_clusters_angular_positions_Human5}
Neuroanatomical clusters of Human5 superimposed over the inferred positions in the hyperbolic embeddings.  The distribution of the angular positions is also shown, as well as the ratio $\langle \Delta \theta \rangle^\mathrm{cluster} / \langle \Delta \theta \rangle^\mathrm{rand.}$, where $\langle \Delta \theta \rangle^\mathrm{cluster}$ corresponds to the average angular separation between nodes of a same cluster, and where $\langle \Delta \theta \rangle^\mathrm{rand.}$ is the average angular separation between the nodes of a cluster of the same size but composed of nodes chosen at random (10000 samples simulated to compute $\langle \Delta \theta \rangle^\mathrm{rand.}$).  Also shown is an estimate of the probability $P$ for these random clusters to have an average angular separation equal to smaller than $\langle \Delta \theta \rangle^\mathrm{cluster}$ (i.e. estimated by the fraction of the 10000 randomly generated samples).  The same information, but computed using the distances in Euclidean space, is also shown for comparison.

\paragraph*{S7 Fig.}
\label{fig:anatomical_clusters_angular_positions_Human3}
Neuroanatomical clusters of Human3 superimposed over the inferred positions in the hyperbolic embeddings.  The distribution of the angular positions is also shown, as well as the ratio $\langle \Delta \theta \rangle^\mathrm{cluster} / \langle \Delta \theta \rangle^\mathrm{rand.}$, where $\langle \Delta \theta \rangle^\mathrm{cluster}$ corresponds to the average angular separation between nodes of a same cluster, and where $\langle \Delta \theta \rangle^\mathrm{rand.}$ is the average angular separation between the nodes of a cluster of the same size but composed of nodes chosen at random (10000 samples simulated to compute $\langle \Delta \theta \rangle^\mathrm{rand.}$).  Also shown is an estimate of the probability $P$ for these random clusters to have an average angular separation equal to smaller than $\langle \Delta \theta \rangle^\mathrm{cluster}$ (i.e. estimated by the fraction of the 10000 randomly generated samples).  The same information, but computed using the distances in Euclidean space, is also shown for comparison.

\paragraph*{S8 Fig.}
\label{fig:anatomical_clusters_angular_positions_Human2}
Neuroanatomical clusters of Human2 superimposed over the inferred positions in the hyperbolic embeddings.  The distribution of the angular positions is also shown, as well as the ratio $\langle \Delta \theta \rangle^\mathrm{cluster} / \langle \Delta \theta \rangle^\mathrm{rand.}$, where $\langle \Delta \theta \rangle^\mathrm{cluster}$ corresponds to the average angular separation between nodes of a same cluster, and where $\langle \Delta \theta \rangle^\mathrm{rand.}$ is the average angular separation between the nodes of a cluster of the same size but composed of nodes chosen at random (10000 samples simulated to compute $\langle \Delta \theta \rangle^\mathrm{rand.}$).  Also shown is an estimate of the probability $P$ for these random clusters to have an average angular separation equal to smaller than $\langle \Delta \theta \rangle^\mathrm{cluster}$ (i.e. estimated by the fraction of the 10000 randomly generated samples).  The same information, but computed using the distances in Euclidean space, is also shown for comparison.

\paragraph*{S9 Fig.}
\label{fig:anatomical_clusters_angular_positions_Human1}
Neuroanatomical clusters of Human1 superimposed over the inferred positions in the hyperbolic embeddings.  The distribution of the angular positions is also shown, as well as the ratio $\langle \Delta \theta \rangle^\mathrm{cluster} / \langle \Delta \theta \rangle^\mathrm{rand.}$, where $\langle \Delta \theta \rangle^\mathrm{cluster}$ corresponds to the average angular separation between nodes of a same cluster, and where $\langle \Delta \theta \rangle^\mathrm{rand.}$ is the average angular separation between the nodes of a cluster of the same size but composed of nodes chosen at random (10000 samples simulated to compute $\langle \Delta \theta \rangle^\mathrm{rand.}$).  Also shown is an estimate of the probability $P$ for these random clusters to have an average angular separation equal to smaller than $\langle \Delta \theta \rangle^\mathrm{cluster}$ (i.e. estimated by the fraction of the 10000 randomly generated samples).  The same information, but computed using the distances in Euclidean space, is also shown for comparison.

\paragraph*{S10 Fig.}
\label{fig:anatomical_clusters_angular_positions_Human8}
Neuroanatomical (lobes and gyri levels identified in Ref.~\cite{Fan:2016}) and functional clusters (7 and 17 cluster parcellation identified in Ref.~\cite{Yeo2011}) of Human8 superimposed over the inferred positions in the hyperbolic embeddings. The distribution of the angular positions is also shown, as well as the ratio $\langle \Delta \theta \rangle^\mathrm{cluster} / \langle \Delta \theta \rangle^\mathrm{rand.}$, where $\langle \Delta \theta \rangle^\mathrm{cluster}$ corresponds to the average angular separation between nodes of a same cluster, and where $\langle \Delta \theta \rangle^\mathrm{rand.}$ is the average angular separation between the nodes of a cluster of the same size but composed of nodes chosen at random (10000 samples simulated to compute $\langle \Delta \theta \rangle^\mathrm{rand.}$).  Also shown is an estimate of the probability $P$ for these random clusters to have an average angular separation equal to smaller than $\langle \Delta \theta \rangle^\mathrm{cluster}$ (i.e. estimated by the fraction of the 10000 randomly generated samples).  The same information, but computed using the distances in Euclidean space, is also shown for comparison.





\section*{Acknowledgments}
We would like to thank the authors who shared their connectome data either publicly or privately. We would also like to thank Mari\'an Bogu\~n\'a, Guillermo Garc\'ia-P\'erez, Francesco A. Massucci.


%
%
%





\includepdf[pages=-,fitpaper]{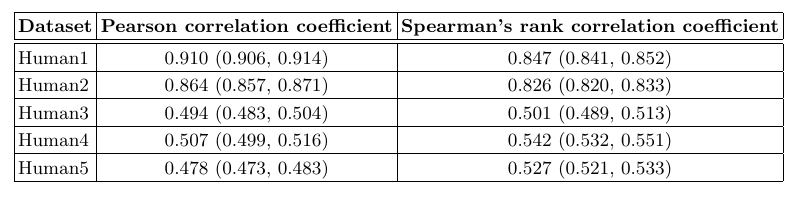}
\includepdf[pages=-,fitpaper]{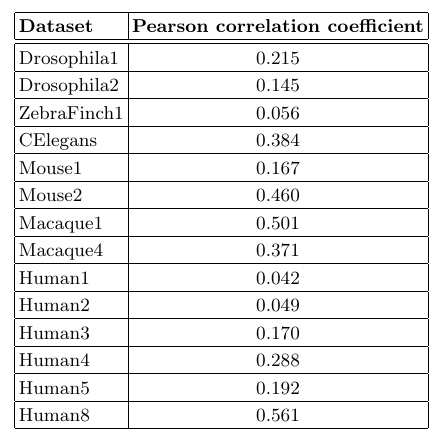}

\includepdf[pages=-,fitpaper]{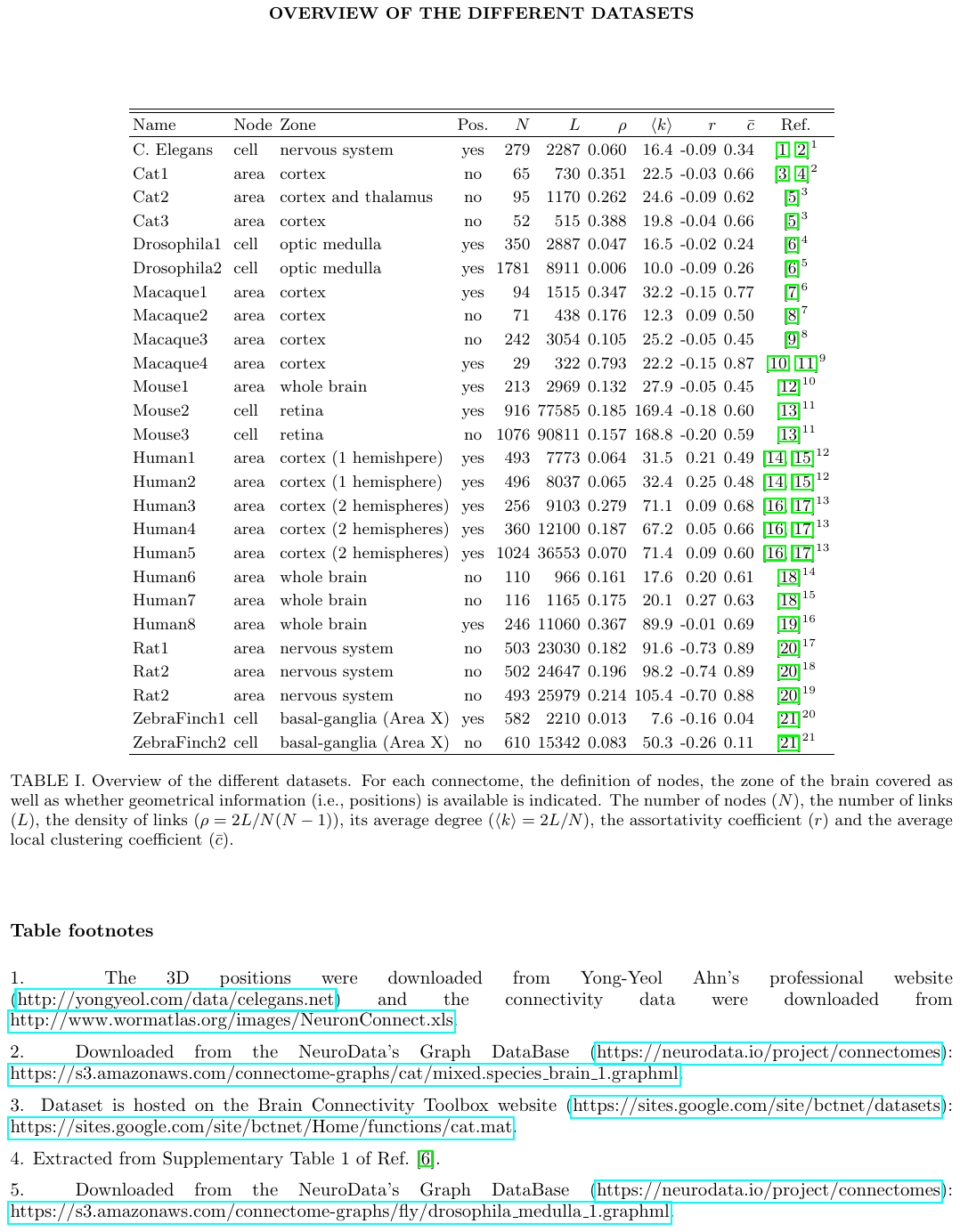}
\includepdf[pages=-,fitpaper]{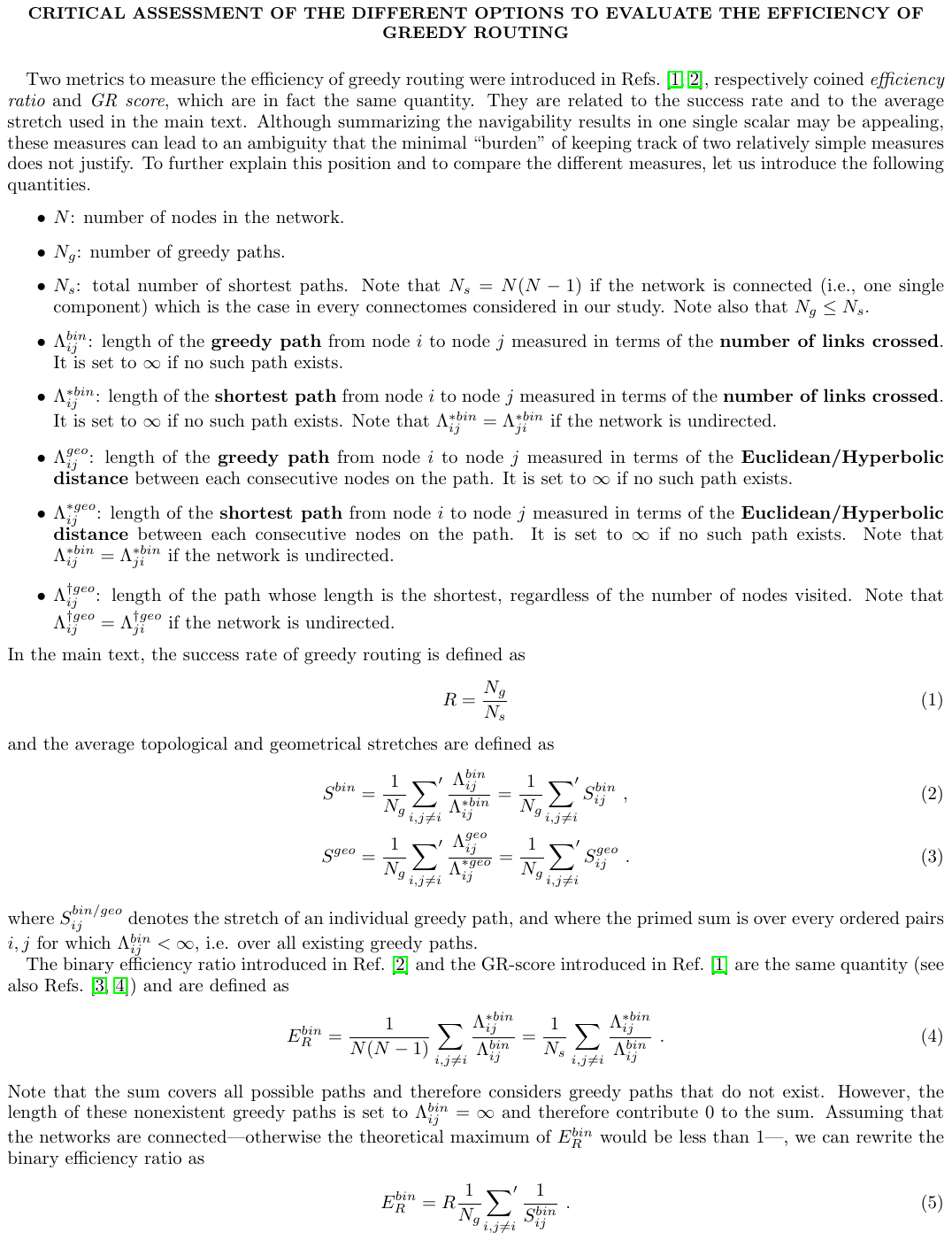}
\includepdf[pages=-,fitpaper]{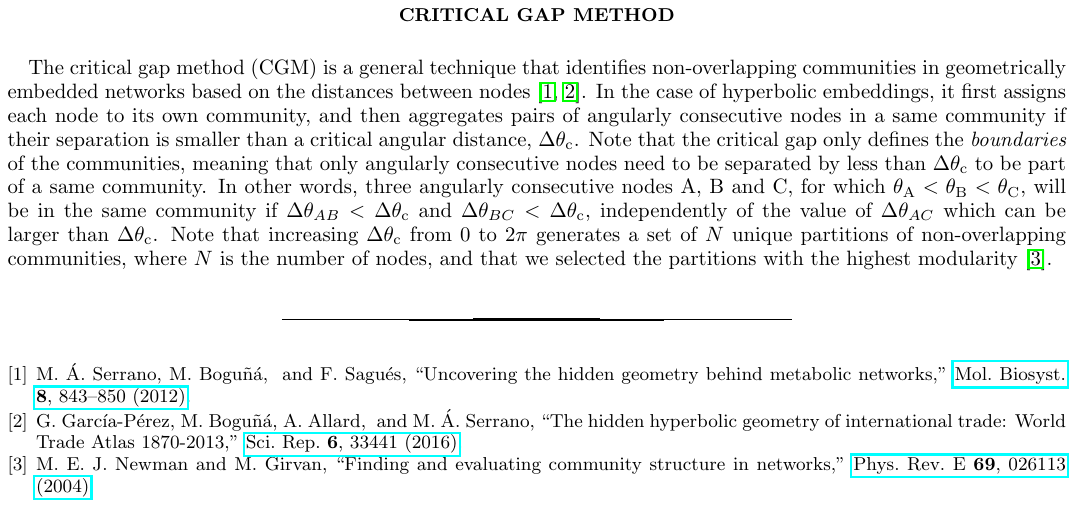}

\includepdf[pages=-,fitpaper]{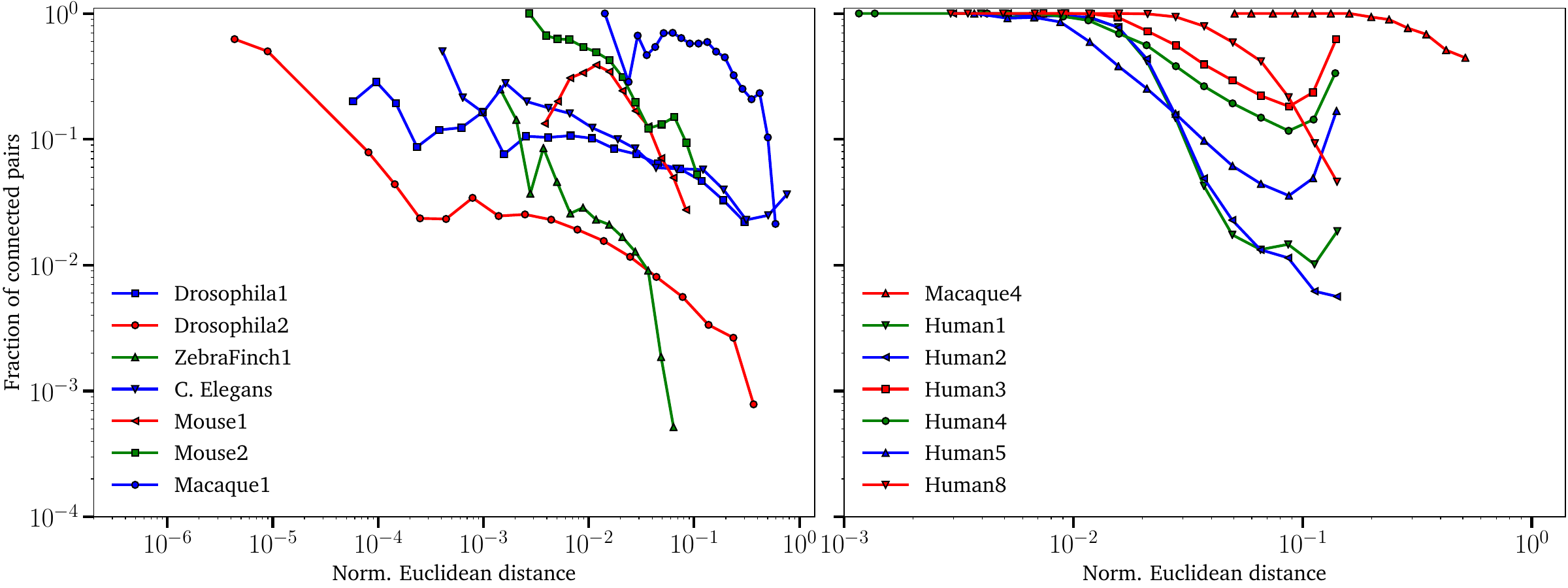}
\includepdf[pages=-,fitpaper]{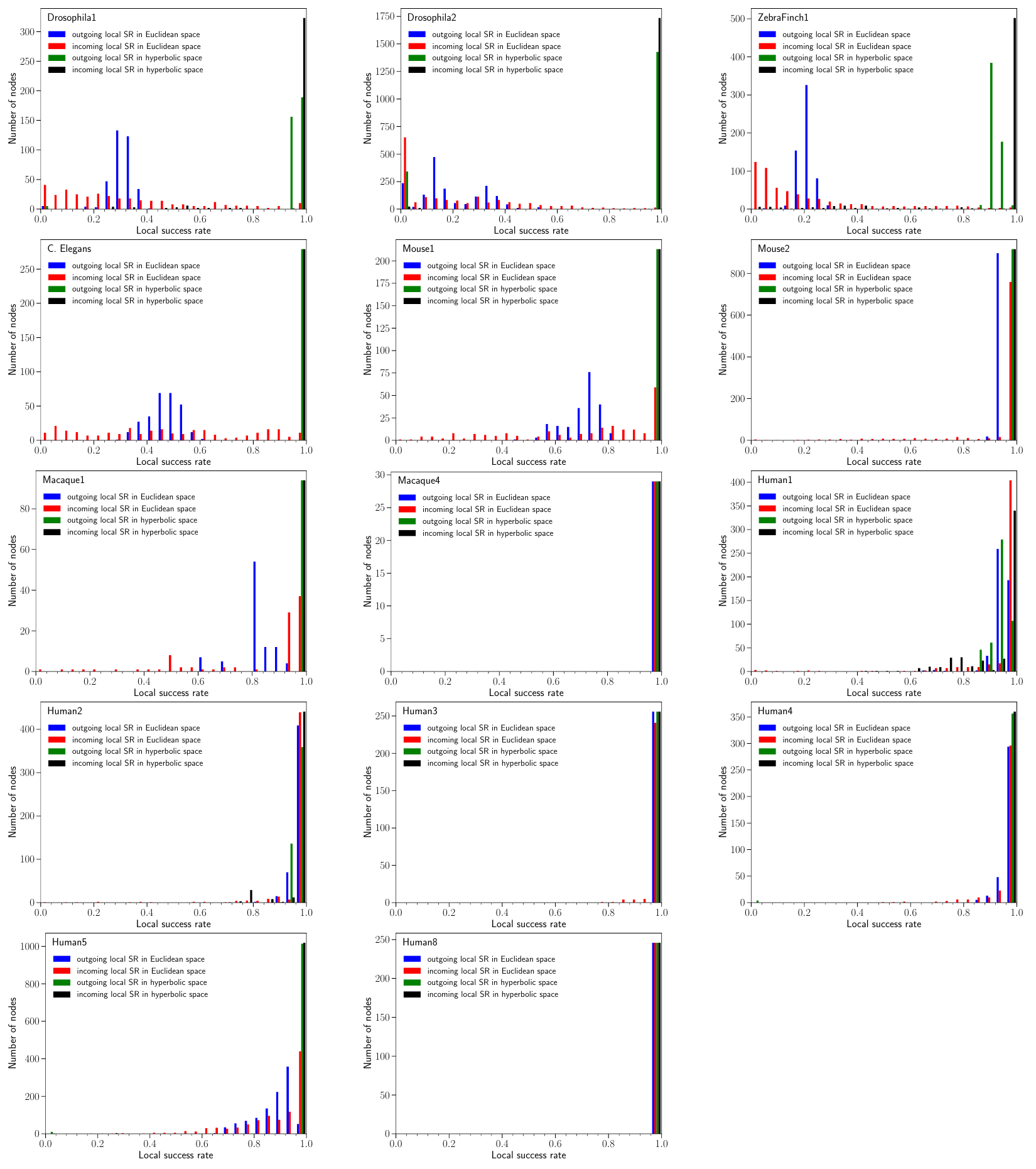}
\includepdf[pages=-,fitpaper]{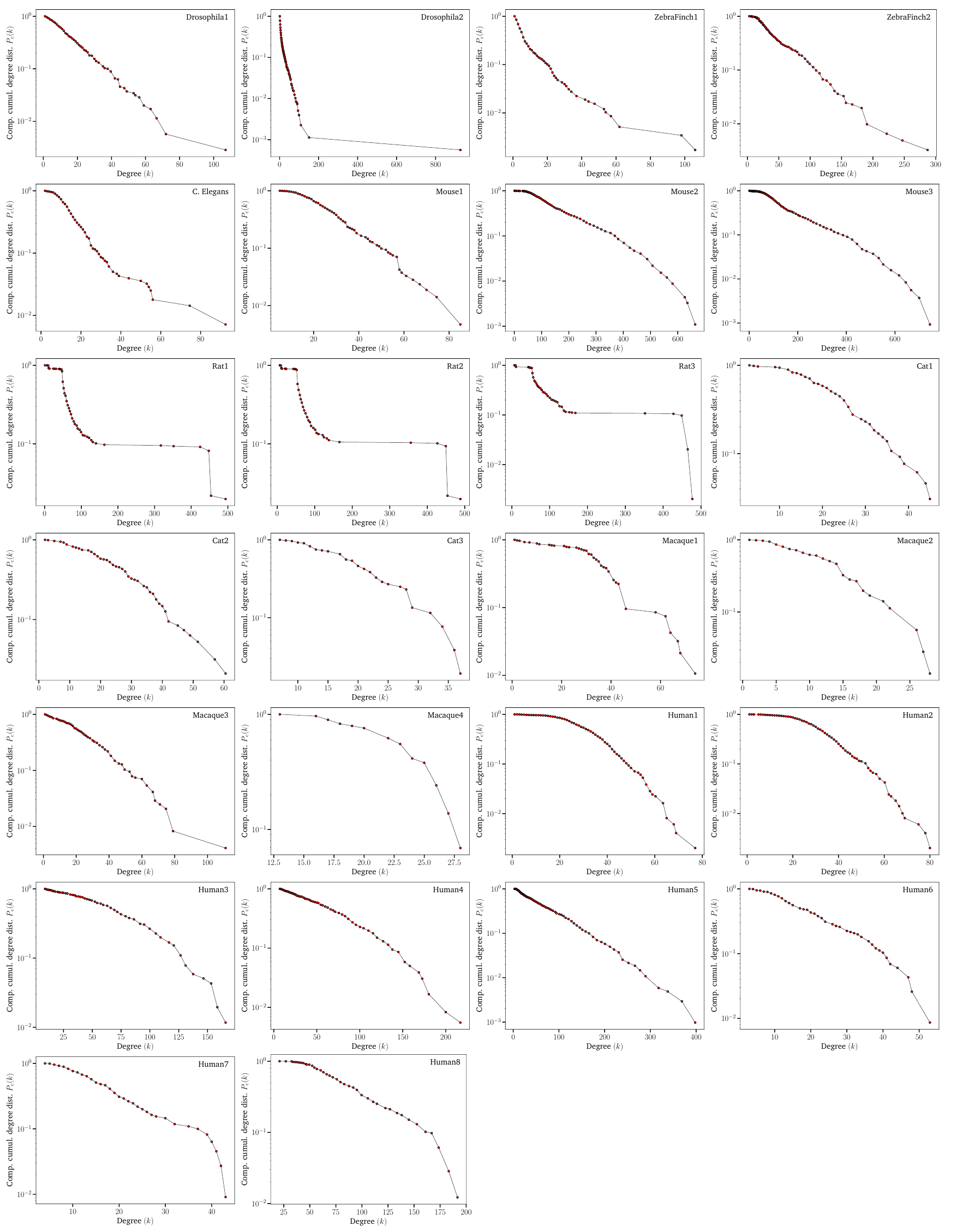}
\includepdf[pages=-,fitpaper]{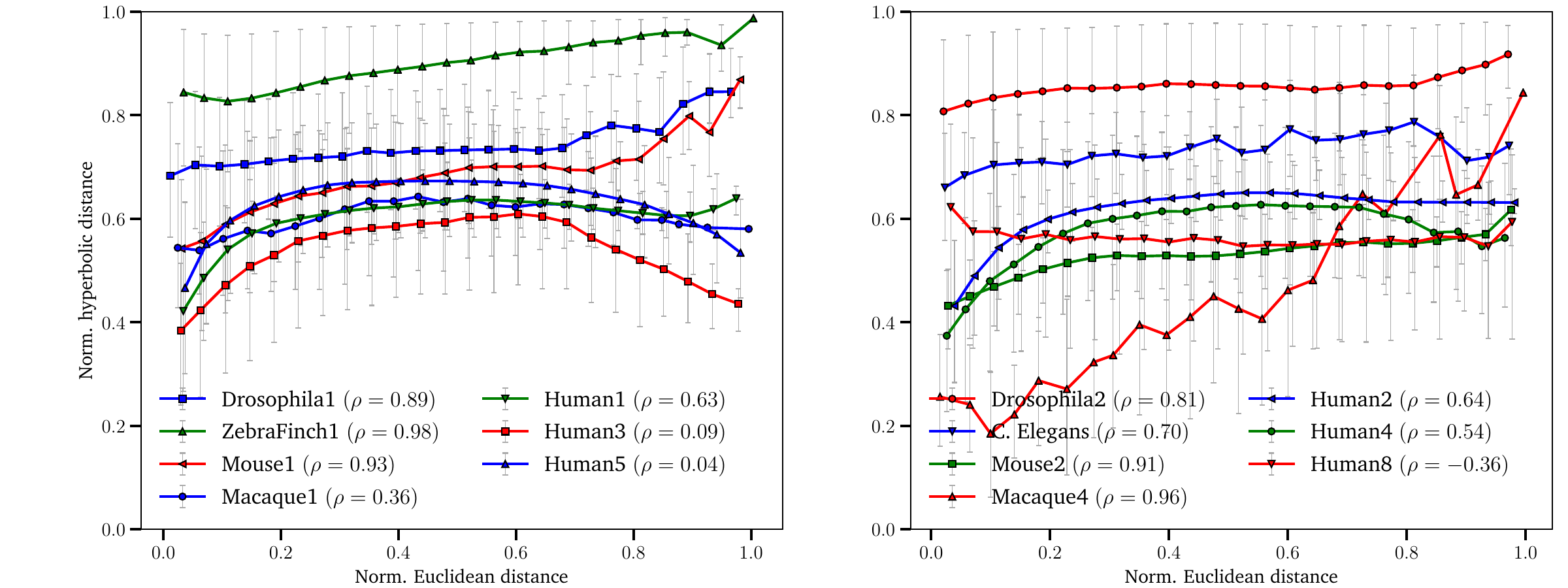}
\includepdf[pages=-,fitpaper]{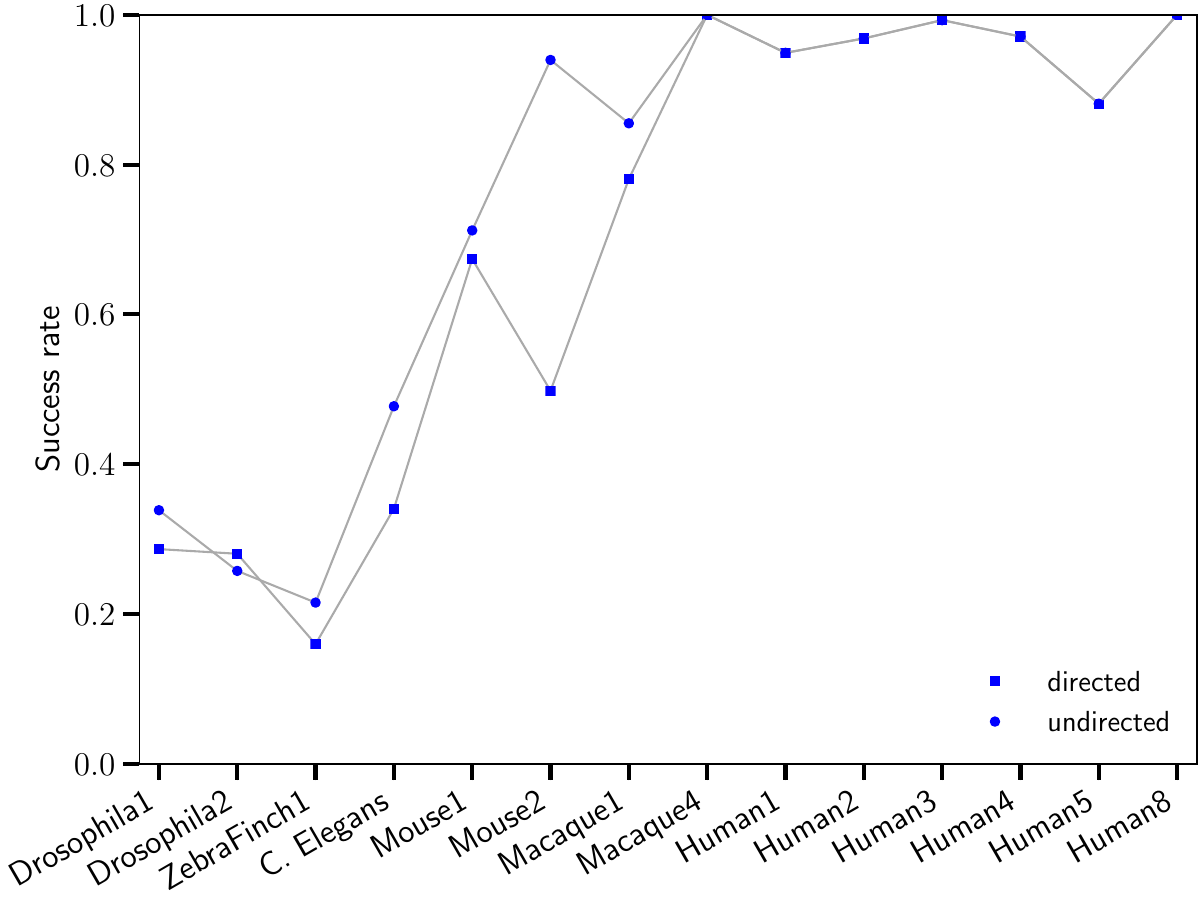}
\includepdf[pages=-,fitpaper]{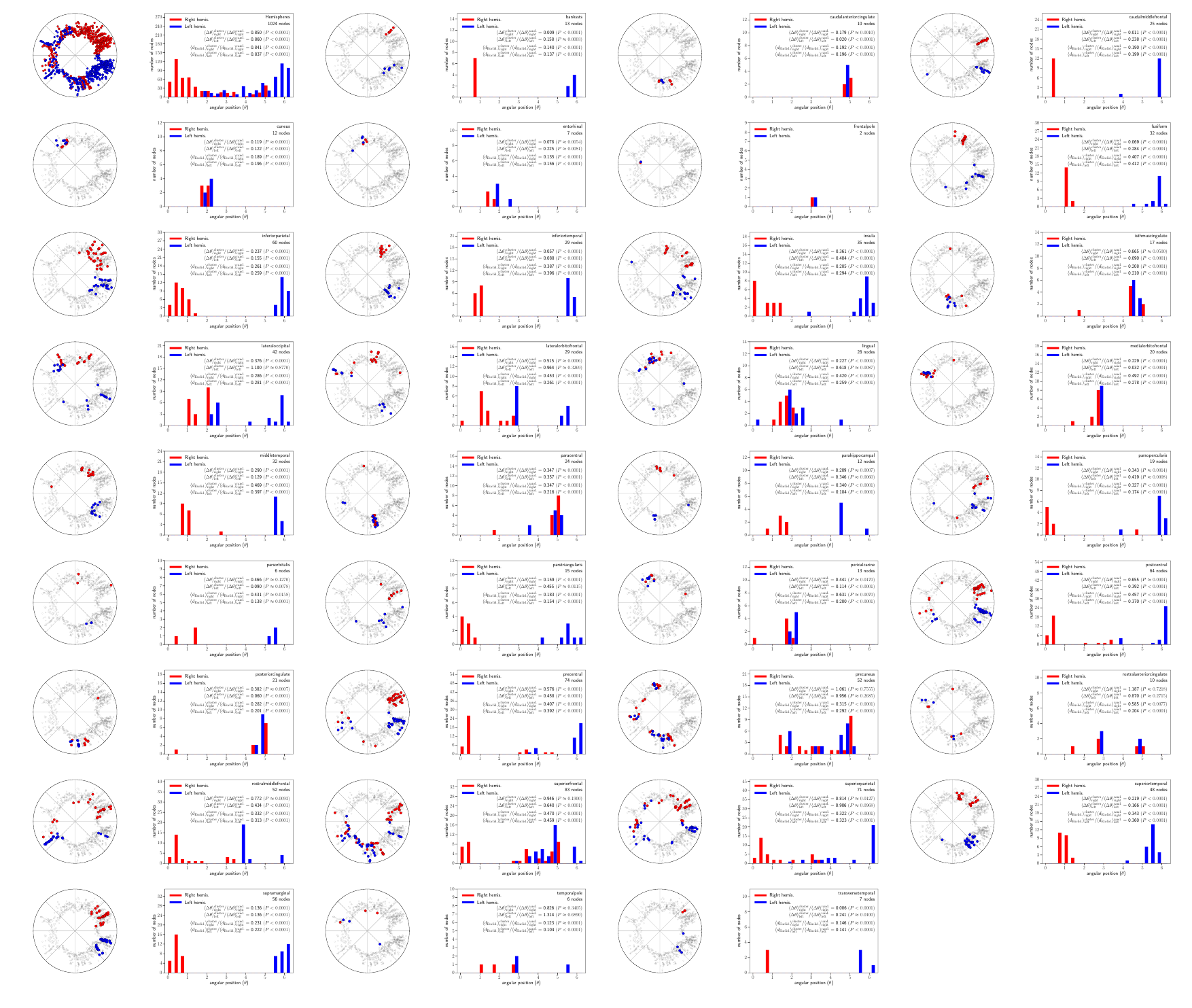}
\includepdf[pages=-,fitpaper]{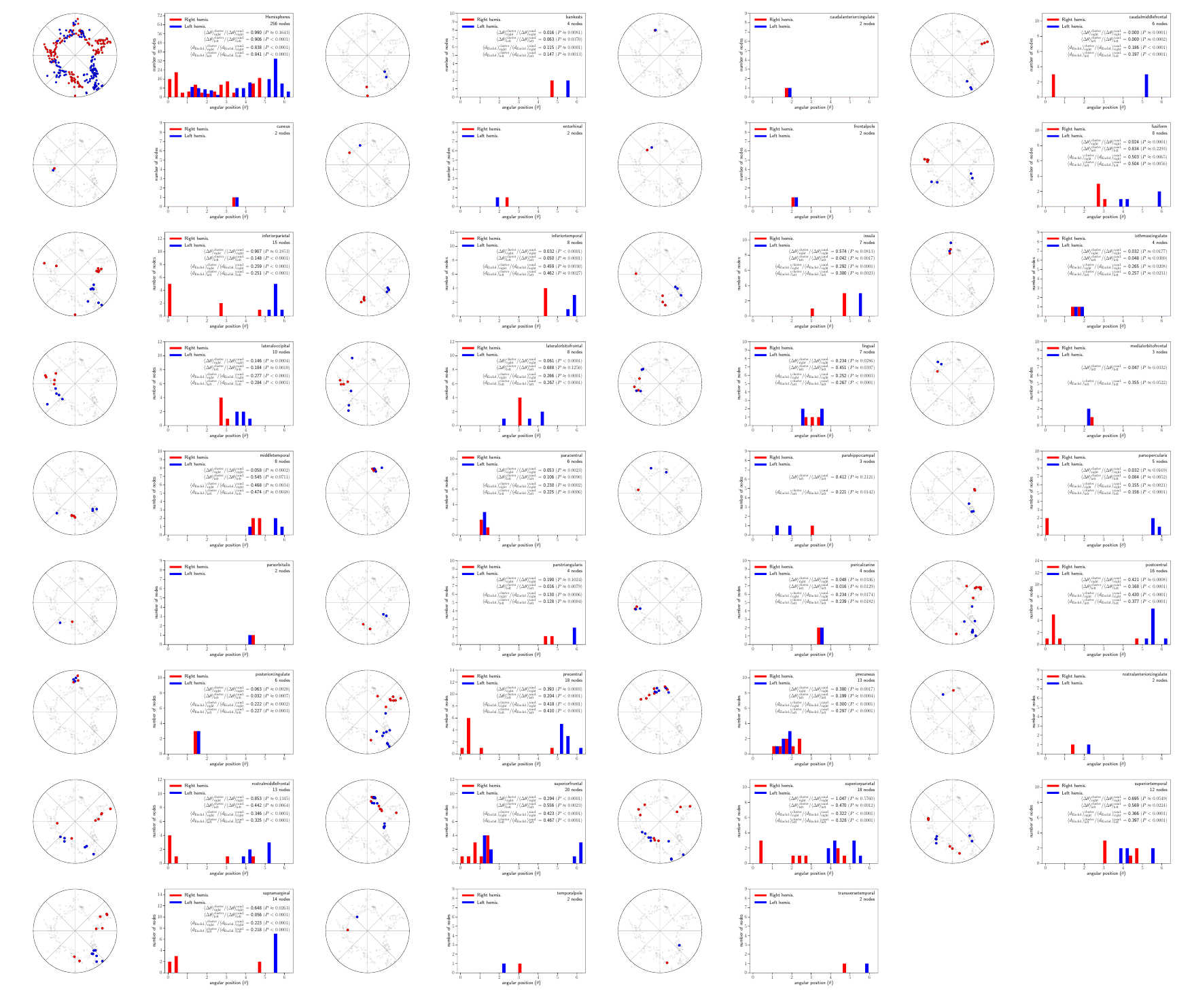}
\includepdf[pages=-,fitpaper]{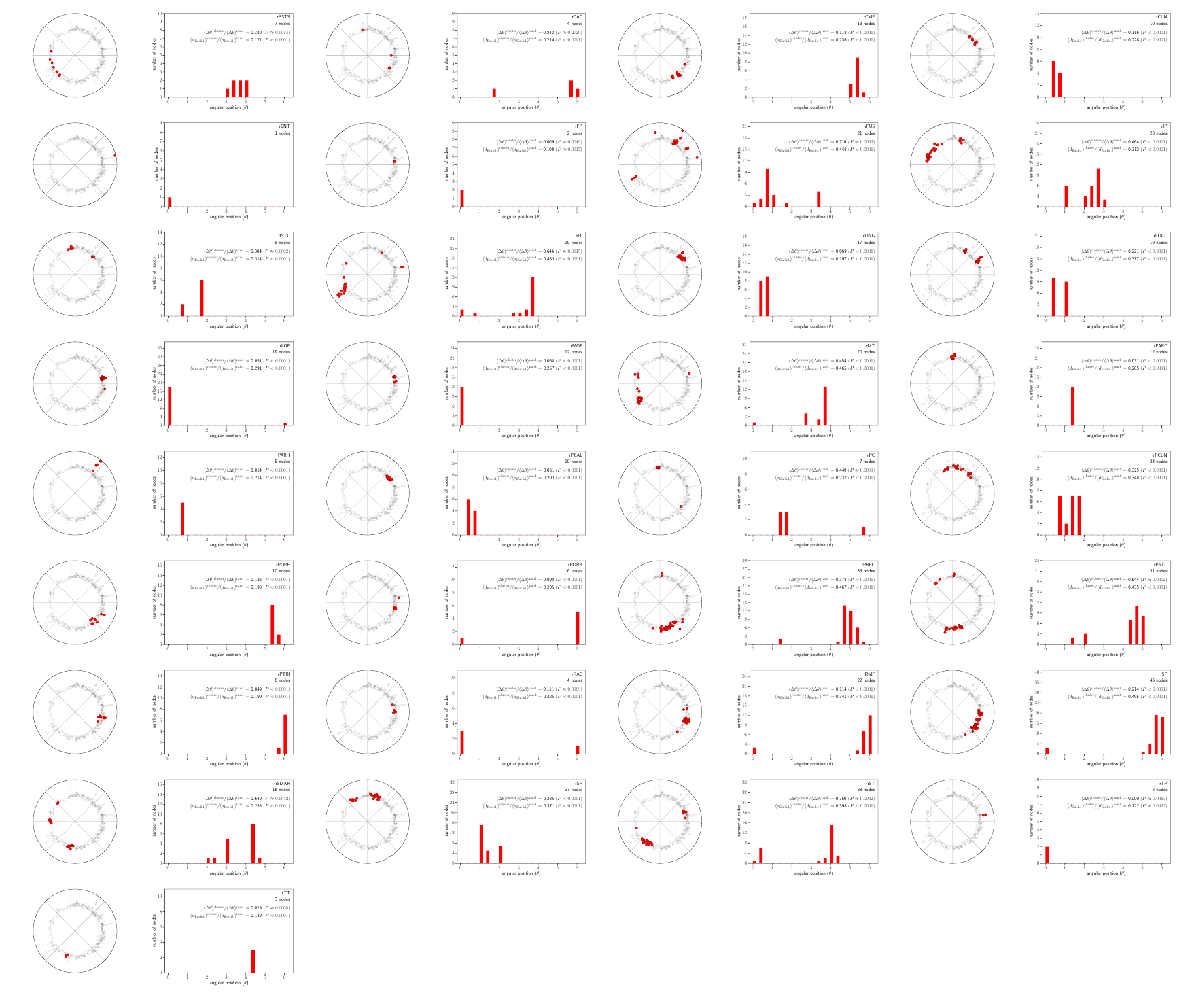}
\includepdf[pages=-,fitpaper]{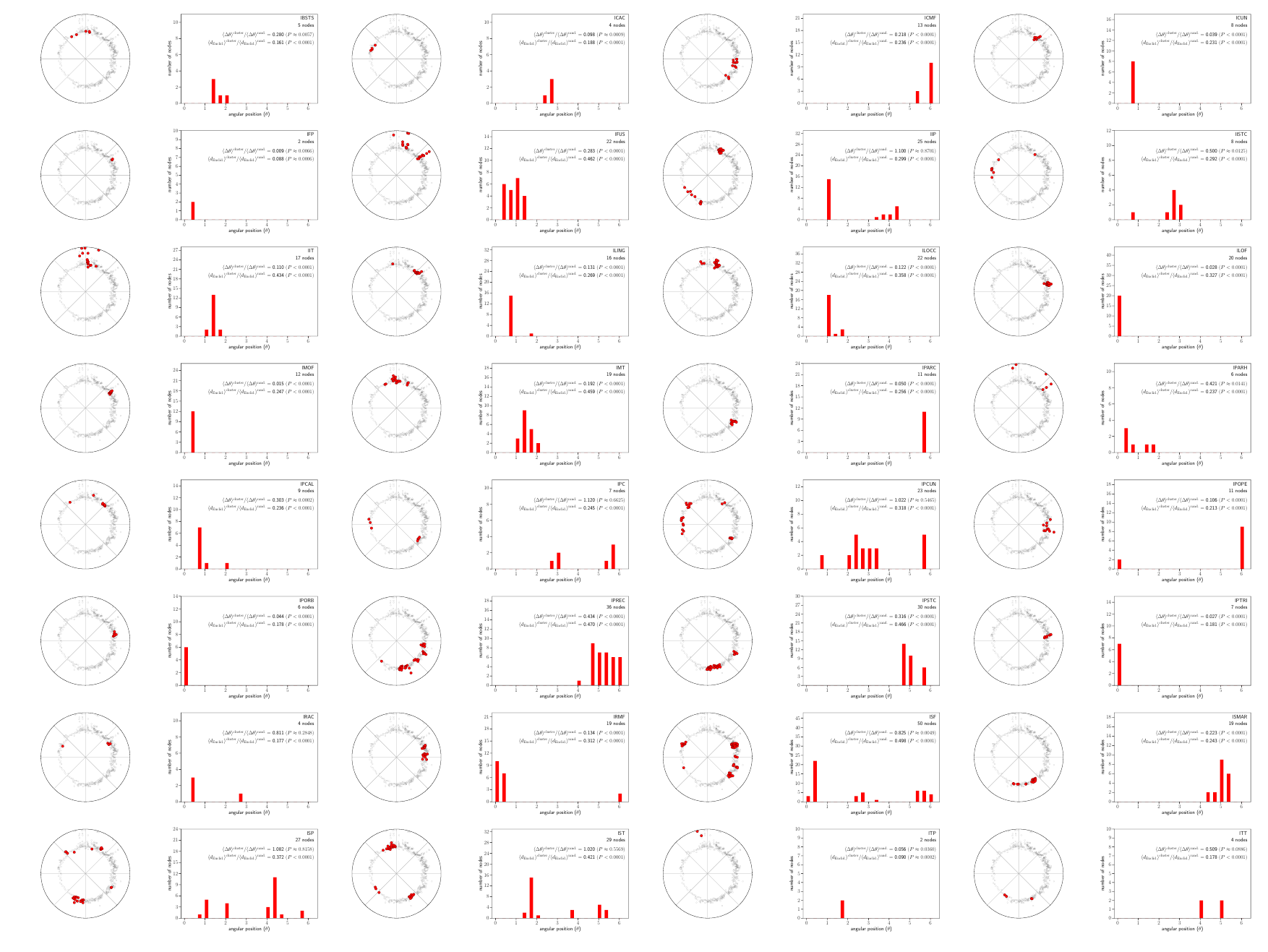}
\includepdf[pages=-,fitpaper]{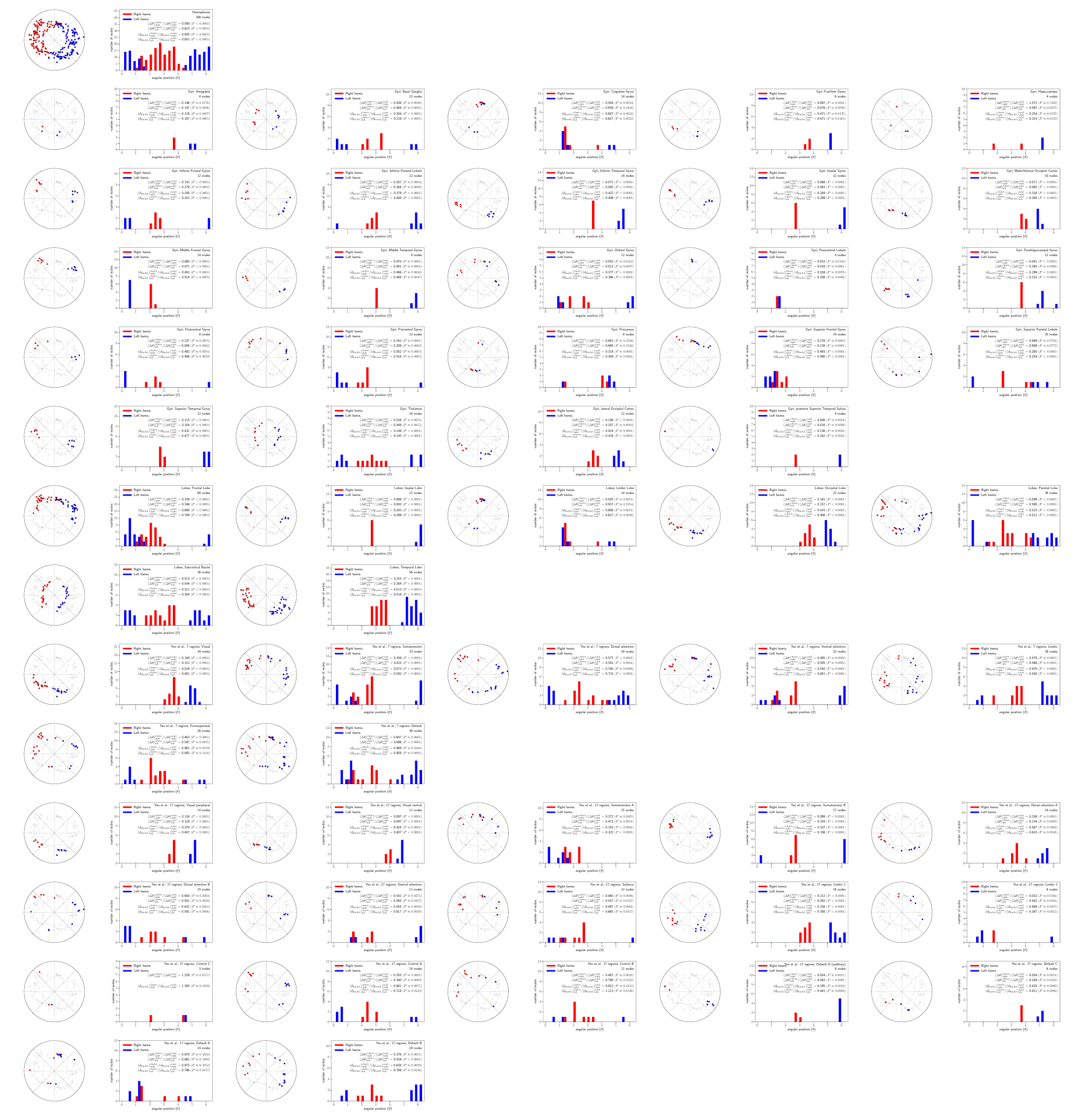}

\end{document}